\documentclass[journal,twocolumn]{IEEEtran}
\usepackage{cite}
\usepackage{cuted}
\usepackage{amsmath,amssymb,amsfonts}
\usepackage{cases}
\usepackage{multirow}
\usepackage{makecell}
\usepackage{booktabs}
\usepackage{algorithmic}
\usepackage{graphicx}
\usepackage{textcomp}
\usepackage{xpatch}
\usepackage{color, xcolor, soul, framed}

\begin{document}
\title{\bf A SVD-based Dynamic Harmonic Phasor Estimator with Improved Suppression of Out-of-Band Interference}

\author{Dongfang Zhao, Shisong Li,~\IEEEmembership{Senior Member,~IEEE}, Fuping Wang, Wei Zhao, \\Songling Huang,~\IEEEmembership{Senior Member,~IEEE}, Qing Wang,~\IEEEmembership{Senior Member,~IEEE}
\thanks{
D. Zhao, S. Li, F. Wang, W. Zhao, and S. Huang are with the Department of Electrical Engineering, Tsinghua University, Beijing 100084, China. Q. Wang is with the Department of Engineering, Durham University, Durham DH1 3LE, UK.
}
}

\maketitle

\begin{abstract}
The diffusion of nonlinear loads and power electronic devices in power systems deteriorates the signal environment, and increases the difficulty of measuring harmonic phasors. Considering accurate harmonic phasor information is necessary to deal with harmonic-related issues, this paper focuses on realizing accurate dynamic harmonic phasor estimation when the signal is contaminated by certain interharmonic tones, i.e. the out-of-band interference (OBI). Specifically, this work introduces the singular value decomposition into the Taylor-Fourier transform algorithm, deriving a general decomposition form for harmonic phasor filters, and yielding a set of adjustable parameters. Then these parameters are configured to minimize the negative influence of OBI on the dynamic harmonic phasor filters. Based on the recommended parameter values, the optimized harmonic estimator is obtained, and then tested under signals distorted by OBI, harmonic, noise, and in worse cases, frequency deviation, and some dynamic conditions. Test results show the proposal achieves good harmonic phasor estimation even under multiple interferences and dynamic scenarios, and has a much weaker dependence on OBI tones compared to conventional approaches.
\end{abstract}

\IEEEpeerreviewmaketitle

\section{Introduction}
\IEEEPARstart{T}{he} power system signals are increasingly distorted by the harmonics and interharmonics arising from the proliferation of nonlinear loads and power electronics-based devices \cite{jain2016fast, cao2017hadoop}. It is necessary to measure harmonic phasors for two folder reasons. First, it is a key step to solving problems related to harmonics, such as harmonic state estimation, harmonic source localization, harmonic active filter design, and power quality analysis \cite{blanco2016implementation}. Second, the harmonic phasor estimation is of great significance to excite some potential applications. For example, precise harmonic phasor estimation is beneficial to improve the accuracy of identifying distribution topology \cite{chen2020switch}; the development of large power system simulations needs the joint effort of accurate and fast harmonic phasor estimators, and electromagnetic transient simulators \cite{rupasinghe2021assessment}; the dynamic harmonic phasor can be used for small-signal stability analysis and fast numerical simulation of microgrid \cite{peng2020modeling}; accurate harmonic information can also be used in locating high impedance fault \cite{farajollahi2017location}, and arcing fault \cite{lin2004new}. However, the dynamic harmonic parameter change, and undesired interharmonic interference bring great difficulties to the accurate and fast estimation of harmonic phasors.

The discrete-time transform (DFT) is an easy and fast algorithm to extract the steady harmonic phasors. However, when dealing with the time-variant harmonics caused by load variations, the DFT accuracy will suffer from the spectrum leakage and fence effect \cite{ferrero2016dynamic, de2015identification}. Subsequently, improved frequency-domain algorithms based on DFT have been employed for harmonic phasor estimation. The IEC 61000-4-7 standard employed the grouping harmonic to get the harmonic amplitude, but the phase estimation was unavailable \cite{emc2002}. Besides, compressive sensing algorithm is beneficial to accurate harmonic phasor estimation by enhancing the frequency resolution \cite{reddy2019open}. However, the improvement is limited, and the algorithm has the disadvantage of a considerable computation burden increase.

To realize accurate harmonic phasor estimation under dynamic conditions, some time-domain algorithms try to widen the passband width of harmonic phasor filters. For example, the Taylor-Fourier transform (TFT) \cite{platas2010dynamic}, and O-Splines \cite{de2020dynamic} employed the Taylor series to extract dynamic harmonic phasors. The limitation, however, is that the passband widths of harmonic filters are identical to each harmonic, diverging from being ideally proportional to harmonic orders. Thus, they are less accurate for estimating high order harmonic phasors than low orders.

Considering the above concerns, \cite{chen2018dynamic, chen2020harmonic} respectively presented the sinc interpolation function, and frequency-domain sampling theorem to replace the Taylor basis function. These approaches allow realizing dynamic harmonic phasor filters (DHPFs) with passband width positively related to harmonic order, and hence can yield a better estimation under frequency deviation and time-variant conditions. However, they become less accurate when dealing with signals contaminated by the out-of-band interference (OBI), e.g. 330\,Hz for six-order harmonic in an electric power system \cite{qian2007interharmonics}. In this paper, the OBI emphasizes the interharmonic interference in a specific frequency range that produces negative effect to harmonic phasor transmission \cite{relays2018118}, and the frequency range is named as transition band for harmonic filters. Increasing the time window length can enable these time-domain algorithms to suppress OBI at the cost of reducing response speed.

In order to suppress interharmonics within a limited window length, the commonly adopted practice is to design filter notches at the interference frequencies. The first step is to identify and track the real-time change of harmonic and interharmonic frequencies. Some frequency tracking approaches can be used to achieve this step, such as the Prony \cite{qi2009prony}, matrix pencil (MP) \cite{sheshyekani2016general}, and ESPRIT \cite{jain2016fast} algorithms. The deduced signal frequencies are then incorporated into the signal model. Finally, the least-square algorithm is employed to realize filters with wide passband at desired signal frequencies, and notches at all other interference frequencies. However, these algorithms realize good dynamic performance and suppression of OBI at the cost of a heavy computational burden, and the effect will be compromised when the identified signal component frequencies are not accurate.

As mentioned, when the Taylor signal model is changed to sinc interpolation function \cite{chen2018dynamic}, or frequency-domain sampling theorem \cite{chen2020harmonic}, the dynamic performance of the harmonic filter can be improved. Similarly, it can be deduced that changing the matrix composed of signal model samples should also be able to influence the filter performance. Moreover, our earlier work shows that singular value decomposition (SVD) is indeed helpful to design a synchrophasor estimator with good suppression of OBI \cite{zhao2021svd}. However, it is worth noting that estimating harmonic phasors is more difficult than synchrophasor. Because the higher the harmonic order is, the wider the passband is required. Besides, there may be fundamental leakage interference, harmonic mutual interference, and multiple OBI components. Therefore, the proposed SVD-based dynamic harmonic phasor estimator (SVDHPE) is a step ahead of the research activity proposed in \cite{zhao2021svd}. The main contributions of this paper include:
\begin{itemize}
	\item The proposal realizes good suppression of the OBI tones when dealing with steady and dynamic signals within a limited window (three nominal cycles), and without need to know signal frequencies in advance.
	\item In this paper, we introduce the singular value decomposition into the Taylor-Fourier transform algorithm, and derive a general decomposition form for harmonic filters obtained by TFT.
	\item Thanks to SVD, we present some adjustable parameters, and then construct an optimization problem about the transition band of harmonic phasor filters.
	\item We fully discuss and evaluate the proposed SVDHPE algorithm from different time window lengths, Taylor expansion orders, numerical and experimental test signals, and the required computational burden.
\end{itemize}

\section{Harmonic Phasor Estimator Design based on Singular Value Decomposition}
\label{sec2}
\subsection{Introducing the SVD into the TFT algorithm}
\label{sec2A}
This section recalls the dynamic harmonic phasor filters designed by the TFT algorithm \cite{platas2010dynamic}, and induces how to introduce the SVD into the harmonic filters' formula. To start with, the power system voltage or current signal distorted by harmonics can be modeled as
\begin{align}
    s(t)=\sum\limits_{h=1}^{H}A_h(t)\cos(2\pi hf_\text{r}t+\phi_h(t))=\sum\limits_{h=1}^{H}s_h(t), 
    \label{eq:1}
\end{align}
where $H$ is the maximum harmonic order; $f_\text{r}$ is the fundamental frequency; $A_h(t)$ and $\phi_h(t)$ respectively denote the amplitude and phase of $s_h(t)$, i.e., fundamental ($h=1$) or harmonic ($h\in[2,\,H]$), and the two elements can be combined to the dynamic phasor $p_h(t)=A_h(t)\text{e}^{\text{j}\phi_h(t)}$. Further, the signal $s(t)$ is sampled at frequency $f_\text{s}$, and then each component $s_h(t+nT_\text{s})$ ($T_\text{s}=1/f_\text{s}$ is the sampling interval) can be written by its Taylor series expansion $p_h(t+\tau)\approx\sum\limits_{k=0}^{K}\frac{\tau^k}{k!}p_{h,k}(t)$ \cite{platas2010dynamic}, i.e.,
\begin{align}
		s_h(t+nT_\text{s})=&\sum\limits_{k=0}^{K}\frac{(nT_\text{s})^k}{k!}0.5p_{h,k}(t)\text{e}^{\text{j}2\pi hf_\text{r}(t+nT_\text{s})}\nonumber\\
		&+\sum\limits_{k=0}^{K}\frac{(nT_\text{s})^k}{k!}0.5p^*_{h,k}(t)\text{e}^{\text{-j}2\pi hf_\text{r}(t+nT_\text{s})}\nonumber\\
		=&\text{e}^{\text{j}2\pi hf_\text{r}nT_\text{s}}\boldsymbol{b}_n\boldsymbol{p}_{h}+\text{e}^{\text{-j}2\pi hf_\text{r}nT_\text{s}}\boldsymbol{b}_n\boldsymbol{p}^*_{h},
		\label{eq:2}
\end{align}
where $K$ is the maximum order of the Taylor series; $^*$ denotes the conjugate operation of a complex number; row vector $\boldsymbol{b}_n=[1,\,\cdots,\,(nT_\text{s})^K/K!]$; column vector $\boldsymbol{p}_{h}=[0.5p_{h,0}(t)\text{e}^{\text{j}2\pi hf_\text{r}t}, \,\cdots,\,0.5p_{h,K}(t)\text{e}^{\text{j}2\pi hf_\text{r}t}]^{\text{T}}$, and $^\text{T}$ represents the transpose operation of a vector or matrix.

Then, within a time window $T_\text{w}$ where the center time is taken as the time tag, the $N=2N_h+1$ samples for each component yield the following equations
\begin{small}
\begin{align}
	\boldsymbol{S}_h=\boldsymbol{E}_{h}\boldsymbol{B}_{K}\boldsymbol{p}_{h}+\boldsymbol{E}^*_{h}\boldsymbol{B}_{K}\boldsymbol{p}^*_{h}=[\boldsymbol{E}_{h}\,\boldsymbol{E}^*_{h}]
	\left[
	\setlength{\arraycolsep}{0.25pt} 
	\begin{array}{cc}
		\boldsymbol{B}_{K} &\boldsymbol{0}\\
		\boldsymbol{0} &\boldsymbol{B}_{K}
	\end{array}
	\right]
	\left[
	\setlength{\arraycolsep}{0.25pt} 
	\begin{array}{c}
		\boldsymbol{p}_{h}\\
		\boldsymbol{p}^*_{h}
	\end{array}
	\right],
	\label{eq:3}
\end{align}
\end{small}
where $\boldsymbol{S}_h=[s_h(t-N_hT_\text{s}),\,\cdots,\,s_h(t+N_hT_\text{s})]^{\text{T}}$; diagonal $\boldsymbol{E}_h\in\mathbb{C}^{N\times N}$ has diagonal elements as $\text{e}^{\text{j}2\pi hf_\text{r}nT_{\text{s}}}$ ($-N_h\le n\le N_h,\,1\le h\le H$); $\boldsymbol{B}_K\in\mathbb{R}^{N\times(K+1)}$ is composed of $N$ row vectors $\boldsymbol{b}_n$ in $T_\text{w}$; $\boldsymbol{p}_h$ are the Taylor series to be solved.

Eq. (\ref{eq:1}) shows the input data of the TFT algorithm are the sum of $H$ components, and hence the $H$ equations with the form of (\ref{eq:3}) are added together to produce the complete equations as
\begin{small}
\begin{align}
\boldsymbol{S}=
\left[
\setlength{\arraycolsep}{0.27pt} 
\begin{array}{ccccc}
 \boldsymbol{E}_{1}\\ \vdots\\ \boldsymbol{E}_{H}\\\boldsymbol{E}^{*}_{1} \\ \vdots\\ \boldsymbol{E}^{*}_{H}\\ 
\end{array}
\right ]^{\text{T}}
\left[
\setlength{\arraycolsep}{0.27pt} 
\begin{array}{cccccc}
 \boldsymbol{B}_{K}  &        &        &        &        & \boldsymbol{0} \\ 
        & \ddots &        &        &        &        \\ 
        &        & \boldsymbol{B}_{K}  &        &        &  \\ 
        &        &        & \boldsymbol{B}_{K}  &        &   \\
        &        &        &        & \ddots &       \\
 \boldsymbol{0}      &        &        &        &        & \boldsymbol{B}_{K}
\end{array}
\right ]   
\left[
\setlength{\arraycolsep}{0.27pt} 
\begin{array}{cccc}
 \boldsymbol{p}_{1}\\ 
 \vdots \\ 
 \boldsymbol{p}_{H}\\ 
 \boldsymbol{p}^{*}_{1}\\ 
 \vdots \\ 
 \boldsymbol{p}^{*}_{H}\\ 
\end{array}
\right ]
=\boldsymbol{EBP}=\boldsymbol{GP},
\label{eq:4}
\end{align}
\end{small}
where column vector $\boldsymbol{S}\in\mathbb{R}^{N\times 1}$ contains $N$ samples of signal $s(t)$, i.e., $\boldsymbol{S}=[s(t-N_hT_\text{s}),\,\cdots,\,s(t+N_hT_\text{s})]^{\text{T}}$; $\boldsymbol{E}_h$, $\boldsymbol{B}_K$, and $\boldsymbol{p}_h$ are the same as in (\ref{eq:3}).

In (\ref{eq:4}), there are in total $2H(K+1)$ parameters to be solved, and $N_c \times c$ equations, where $N_c=f_\text{s}T_0$ represents the number of samples in one nominal fundamental cycle $T_0$, and $c=T_\text{w}/T_0$ is the number of nominal fundamental cycles in one time window $T_\text{w}$. Note the (\ref{eq:4}) usually gives an overdetermined system of linear equations, i.e. $cN_c\geq 2H(K+1)$, and its optimal solution can be obtained by the least-square algorithm, i.e.,
\begin{equation}
	\hat{\boldsymbol{P}}=((\boldsymbol{EB})^{\text{H}}(\boldsymbol{EB}))^{-1}(\boldsymbol{EB})^{\text{H}}\boldsymbol{S}=\boldsymbol{G}^+\boldsymbol{S},
	\label{eq:5}
\end{equation}
where $^{\text{H}}$, $^{-1}$, and $^+$ respectively denote the conjugate transpose, the inverse, and the generalized inverse operations of a matrix; $\hat{\boldsymbol{X}}$ represents the estimation of $\boldsymbol{X}$. Eq. (\ref{eq:5}) shows that each row elements in the matrix $\boldsymbol{G}^+$ form an finite impulse response (FIR) filter to solve the parameter in the column vector $\boldsymbol{P}$.

It is observed that the matrix $\boldsymbol{B}_K$ in (\ref{eq:4}) is formed by the samples of Taylor series signal model. According to \cite{zhao2021svd}, we employ the SVD to decompose it as (\ref{eq:6}) shows, and derive a general decomposition form in (\ref{eq:10}) for harmonic filters obtained by TFT algorithm.
\begin{equation}
    \boldsymbol{B}_K=\boldsymbol{C\it{\Lambda} D}^\text{T},
    \label{eq:6}
\end{equation}
where the matrix $\boldsymbol{C}\in\mathbb{R}^{N\times (K+1)}$; the orthogonal matrix $\boldsymbol{D}\in\mathbb{R}^{(K+1)\times (K+1)}$; their column vectors $\boldsymbol{c}_i,\,\boldsymbol{d}_i\,(1\le i\le (K+1))$ are left singular vectors and right singular vectors, respectively; $\boldsymbol{\it{\Lambda}} \in\mathbb{R}^{(K+1)\times (K+1)}$ is a diagonal matrix, and its $K+1$ diagonal elements are the singular values.

In (\ref{eq:5}), by replacing its block matrices $\boldsymbol{B}_K$ with the decomposition form in (\ref{eq:6}), the matrix $\boldsymbol{G}^+$ is rewritten as
\begin{align}
    \boldsymbol{G}^{+}=
    \left[
    \setlength{\arraycolsep}{0.15pt}
    \begin{array}{cccccc}
     \boldsymbol{D}\boldsymbol{\it{\Lambda}}^{-1}      &        &        &        &        & \boldsymbol{0} \\ 
            & \ddots &        &        &        &        \\ 
            &        & \boldsymbol{D}\boldsymbol{\it{\Lambda}}^{-1}      &        &        &  \\ 
            &        &        & \boldsymbol{D}\boldsymbol{\it{\Lambda}}^{-1}      &        &   \\
            &        &        &        & \ddots &       \\
     \boldsymbol{0}      &        &        &        &        & \boldsymbol{D}\boldsymbol{\it{\Lambda}}^{-1}
    \end{array}
    \right ]
    \boldsymbol{Q}
    \left[
    \setlength{\arraycolsep}{0.27pt} 
    \begin{array}{c}
    \boldsymbol{C}^{\text{T}}\boldsymbol{E}_{1}^{*} \\ 
     \vdots \\ 
    \boldsymbol{C}^{\text{T}}\boldsymbol{E}_{H}^{*} \\ 
    \boldsymbol{C}^{\text{T}}\boldsymbol{E}_{1} \\
     \vdots \\
    \boldsymbol{C}^{\text{T}}\boldsymbol{E}_{H}
    \end{array}
    \right ],
    \label{eq:7}
\end{align}
where $\boldsymbol{Q}\in\mathbb{C}^{2H(K+1)\times 2H(K+1)}$, and has the following form 
\begin{align}
\resizebox{0.95\hsize}{!}{$\boldsymbol{Q}=
\begin{pmatrix}
\left[
\setlength{\arraycolsep}{0.15pt} 
\begin{array}{cccccc}
 \boldsymbol{C}      &        &        &        &        & \boldsymbol{0} \\ 
        & \ddots &        &        &        &        \\ 
        &        & \boldsymbol{C}      &        &        &  \\ 
        &        &        & \boldsymbol{C}      &        &   \\
        &        &        &        & \ddots &       \\
 \boldsymbol{0}      &        &        &        &        & \boldsymbol{C}
\end{array}
\right ]^{\text{T}}
\left[
\setlength{\arraycolsep}{0.1pt} 
\begin{array}{ccccc}
 \boldsymbol{E}^{*}_{1}\\ \vdots\\ \boldsymbol{E}^{*}_{H}\\\boldsymbol{E}_{1} \\ \vdots\\ \boldsymbol{E}_{H}\\ 
\end{array}
\right ]
\left[
\setlength{\arraycolsep}{0.1pt} 
\begin{array}{ccccc}
 \boldsymbol{E}_{1}\\ \vdots\\ \boldsymbol{E}_{H}\\\boldsymbol{E}^{*}_{1} \\ \vdots\\ \boldsymbol{E}^{*}_{H}\\ 
\end{array}
\right ]^{\text{T}}
\left[
\setlength{\arraycolsep}{0.15pt} 
\begin{array}{cccccc}
 \boldsymbol{C}      &        &        &        &        & \boldsymbol{0} \\ 
        & \ddots &        &        &        &        \\ 
        &        & \boldsymbol{C}      &        &        &  \\ 
        &        &        & \boldsymbol{C}      &        &   \\
        &        &        &        & \ddots &       \\
 \boldsymbol{0}      &        &        &        &        & \boldsymbol{C}
\end{array}
\right ]
\end{pmatrix}^{-1}.$}
\label{eq:8}
\end{align}
To simplify $\boldsymbol{G}^+$, we split $\boldsymbol{Q}$ into $2H\times2H$ block matrices $\boldsymbol{Q}_{h,j}\in\mathbb{C}^{(K+1)\times (K+1)}\,(1\le h,j\le 2H)$. Each $\boldsymbol{Q}_{h,j}$ is formed by the cross elements of row $(h-1)(K+1)+1$ to row $h(K+1)$, and column $(j-1)(K+1)+1$ to column $j(K+1)$. Then, the $K+1$ filters for $h$-order harmonic, i.e. the elements from row $(h-1)(K+1)+1$ to row $h(K+1)$ in $\boldsymbol{G}^+$ can be simplified as 
\begin{align}
    \boldsymbol{G}^+_h=\boldsymbol{D}\boldsymbol{\it{\Lambda}}^{-1}(\sum\limits_{j=1}^{H}\boldsymbol{Q}_{h,j}\boldsymbol{C}^{\text{T}}\boldsymbol{E}^{*}_j+
                        \sum\limits_{j=H+1}^{2H}\boldsymbol{Q}_{h,j}\boldsymbol{C}^{\text{T}}\boldsymbol{E}_{(j-H)}),
\label{eq:9}
\end{align}
where $\boldsymbol{G}^+_h\in\mathbb{C}^{(K+1)\times N}\,(1\le h\le H)$. Note there are $K+1$ row vectors, i.e. $K+1$ filters in the matrix $\boldsymbol{G}^+_h$. Whereas, only the first row vector in $\boldsymbol{G}^+_h$ is related to the topic of this paper, the estimation of harmonic phasor, and the concerned row vector can be expressed in a more detailed form with the elements of matrices $\boldsymbol{D},\,\boldsymbol{\it{\Lambda}},\,\boldsymbol{l}_h$ as
\begin{align}
	 \boldsymbol{r}_h=\sum_{k_{\text{e}}=1}^{K+1}\frac{d_{1,k_{\text{e}}}}{\lambda_{k_{\text{e}}}}\boldsymbol{l}_{k_{\text{e}},:},
	 \label{eq:10}
\end{align} 
where the row vector $\boldsymbol{r}_h\in\mathbb{C}^{1\times N}$ denotes the FIR coefficients of the $h-$order DHPF; $\boldsymbol{l}_{k_{\text{e}},:}\in\mathbb{C}^{1\times N}$ presents the $k_{\text{e}}^\text{th}$ row elements of the matrix $\boldsymbol{l}_h=\sum_{j=1}^{H}\boldsymbol{Q}_{h,j}\boldsymbol{C}^{\text{T}}\boldsymbol{E}^{*}_j+
\sum_{j=(H+1)}^{2H}\boldsymbol{Q}_{h,j}\boldsymbol{C}^{\text{T}}\boldsymbol{E}_{(j-H)}\,(\boldsymbol{l}_h\in\mathbb{C}^{(K+1)\times N})$; $d_{1,k_{\text{e}}}$ and $\lambda_{k_{\text{e}}}$ are the $k_{\text{e}}^\text{th}$ elements in the first row of $\boldsymbol{D}$ and the diagonal of $\boldsymbol{\it{\Lambda}}$, respectively. Then, the dynamic harmonic phasor, amplitude, and phase can be estimated as \cite{platas2010dynamic}
\begin{align}
	\hat{p}_{h,0}(t)=&2\text{e}^{\text{-j}2\pi hf_\text{r}t}\boldsymbol{r}_h\boldsymbol{S},\nonumber\\
    \hat{a}_h(t)=&|\hat{p}_{h,0}(t)|,\nonumber\\
    \hat{\phi}_h(t)=&\angle \hat{p}_{h,0}(t),
    \label{eq:11}
\end{align}
where $|Y|,\,\angle Y$ denotes the magnitude, and phase of complex number $Y$, respectively.

\subsection{Design of dynamic harmonic phasor filter with improved transition band performance}
\label{sec2B}
\begin{figure}[tp!]
	\centering
	\includegraphics[width=0.48\textwidth]{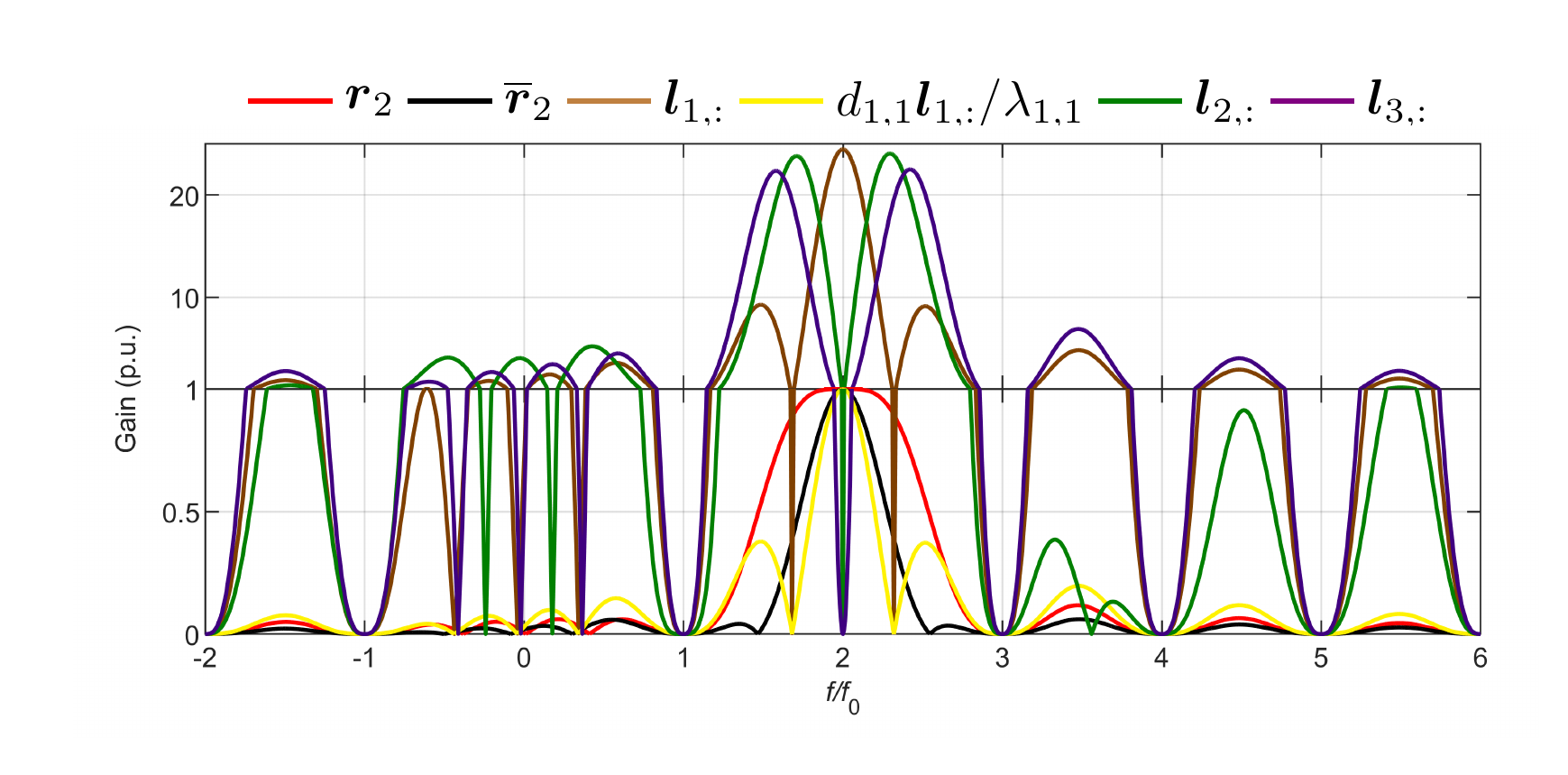}
	\caption{The frequency responses of some filters related to second harmonic.}
	\label{fig:svdtheo}
\end{figure}

Eq. (\ref{eq:10}) shows that each zero-order filter $\boldsymbol{r}_h$ for $h-$order harmonic equals the sum of row elements in $\boldsymbol{l}_h$ weighted by $d_{1,k_{\text{e}}}/\lambda_{k_{\text{e}}}$. An example for $\boldsymbol{r}_2$ when $K=2,\,T_\text{w}=3/f_0$\,($f_0$ is the nominal frequency) is shown in Fig. \ref{fig:svdtheo} to help understand the superposition formulation. It begins with the brown curve $\boldsymbol{l}_{1,:}$ which is not normalized. Then, multiplying $\boldsymbol{l}_{1,:}$ with the weight $d_{1,1}/\lambda_1$ produces the nominalized yellow curve whose passband is narrow. Further, adding the normalized $d_{1,2}\boldsymbol{l}_{2,:}/\lambda_2$ and $d_{1,3}\boldsymbol{l}_{3,:}/\lambda_3$ to $d_{1,1}\boldsymbol{l}_{1,:}/\lambda_1$, the red curve for $\boldsymbol{r}_2$ is obtained, and has more flat gain than the yellow curve. 

The above analysis shows the weights, e.g. $d_{1,1}/\lambda_1$, $d_{1,2}/\lambda_2$, and $d_{1,3}/\lambda_3$ play an important role in forming the filter $\boldsymbol{r}_2$. Also in Fig. \ref{fig:svdtheo}, the black $\overline{\boldsymbol{r}}_2$ is obtained with some easy change to the weights, and it is observed to have lower transition band gain ($f/f_0\in[1,\,1.5]\cup[2.5,\,3]$), and hence higher suppression of interharmonics than the original $\boldsymbol{r}_2$. This inspires us to introduce some multipliers $y_{h,k_{\text{e}}}$ to change the weights $d_{1,k_{\text{e}}}/\lambda_{k_{\text{e}}}$, and realize smaller gain in the OBI frequency range $\Psi_h=[(h-1)f_0,\,hf_0-25]\cup[hf_0+25,\,(h+1)f_0]$, i.e. the transition band for the $h$-order harmonic phasor filter. Then the designed $h-$order harmonic filter is written as
\begin{align}
    \overline{\boldsymbol{r}}_h=\sum_{k_{\text{e}}=1}^{K+1}\frac{d_{1,k_{\text{e}}}}{y_{h,k_{\text{e}}}\cdot\lambda_{k_{\text{e}}}}\boldsymbol{l}_{k_{\text{e}},:},
    \label{eq:12}
\end{align}
where  $y_{h,k_{\text{e}}}$ is the introduced variable to the weight pair $d_{1,k_{\text{e}}}/\lambda_{k_{\text{e}}}$. Namely, $\overline{\boldsymbol{r}}_h$ equals the sum of row elements in $\boldsymbol{l}_h$ weighted by $d_{1,k_{\text{e}}}/(y_{h,k_{\text{e}}}\cdot\lambda_{k_{\text{e}}})$. In case all multipliers $y_{h,k_{\text{e}}}$ equal one, (\ref{eq:12}) gives the original FIR coefficients of the $h-$order DHPF in (\ref{eq:10}), otherwise the FIR coefficients and hence the filtering performance may be altered. 

The following task is to choose suitable $y_{h,k_{\text{e}}}$ from the $K+1$ candidates $(1\le k_{\text{e}}\le K+1)$ to promote the $h-$order DHPF performance. The appendix proves that for the first row elements of the right singular matrix $\boldsymbol{D}$, there is $d_{1,(2a+1)}\neq0$ ($0\le a\le\lfloor K/2\rfloor\text{,}\enskip\lfloor\cdot\rfloor$ represents the round down operation), and $d_{1,(2a)}=0\,(1\le a\le\lfloor(K+1)/2\rfloor)$. Therefore, only the odd multipliers $y_{h,(2a+1)}$ can affect the FIR coefficients, and hence the gain-frequency response of the $h-$order DHPF. 

As a result, we choose the odd multipliers $y_{h,(2a+1)}$ as the adjustable parameters to improve the gain-frequency response of the $h-$order DHPF and, to be simple, set all the even multipliers $y_{h,(2a)}$ to one. However, numerous tests indicate that the improvement in reducing the transition band gain is not infinite. Therefore, we design the following optimization problem aiming to determine the multiplier values that minimize the maximum transition band gain for each harmonic phasor filter, i.e.,
\begin{align}
    \mathop{\textbf{Min}}\limits_{\substack{y_{h,(2a+1)} \\0\le a\le\lfloor K/2\rfloor}}&\{\mathop{\text{max}}\limits_{b\in \Psi_h}(R_{h,b})\}\nonumber\\
    \textbf{s.t.} \quad &\forall \quad y_{h,(2a+1)}>0 \nonumber\\
    &\exists\quad y_{h,(2a+1)}\neq1,
    \label{eq:13}
\end{align}
where max$(R_{h,b})$ denotes the maximum gain of $h-$order harmonic filter when the frequency $b$ covers the transition band $\Psi_h=[(h-1)f_0,\,hf_0-25]\cup[hf_0+25,\,(h+1)f_0]$. The proposal focuses on mitigating the negative effects from interharmonics of frequency $b\in\Psi_h$ when estimating harmonic phasors. It depends on obtaining the optimal $y_{h,(2a+1)}$ to minimize the filter transition band gain by solving (\ref{eq:13}). Once the optimal $y_{h,(2a+1)}$ is obtained and substituted to (\ref{eq:12}), the optimized harmonic filter $\overline{\boldsymbol{r}}_h$ is obtained, and then employed in (\ref{eq:11}) to extract the phasor. Besides, the first restriction is to keep the computation stability. Whereas, the second one denotes that at least one multiplier is changed to alter the filtering performance from the original condition ($y_{h,(2a+1)}=1$).

\begin{figure*}[tp!]
	\centering
	\includegraphics[width=0.95\textwidth]{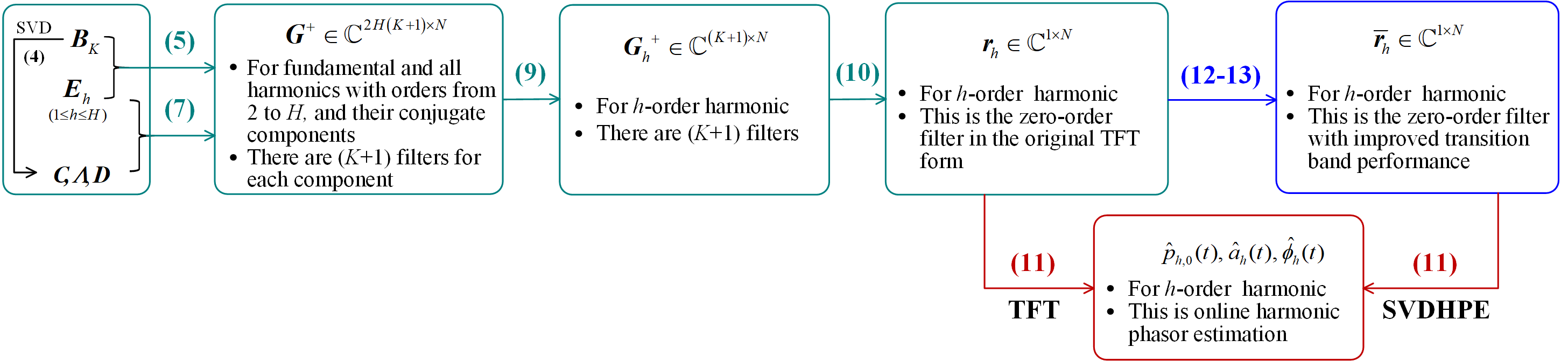}
	\caption{The whole process of the proposed SVDHPE. Eqs. (\ref{eq:5}), (\ref{eq:7}), and (\ref{eq:9})-(\ref{eq:13}) can be found in section \ref{sec2}.}
	\label{fig:conclu}
\end{figure*}

As shown in Fig. \ref{fig:conclu}, the proposed SVDHPE contains three parts. Firstly, the green part shows that we derive a general decomposition form for TFT filters by using SVD. Based on the analysis, the blue part indicates we introduce some multipliers, and propose an optimal problem to improve the transition band performance of DHPFs. The two parts are carried out offline, and aim to obtain the optimized filter coefficients $\overline{\boldsymbol{r}}_h$. Finally, the optimal coefficients $\overline{\boldsymbol{r}}_h$ are used to extract the harmonic phasors online as in the red box, which is simple enough to be deployed into the monitoring devices in substations.

\section{Analysis of Choosing Appropriate Window Length and Taylor Expansion Order}
\label{sec3}
The window length $c$ and Taylor expansion order $K$ are two key parameters to realize the specific TFT harmonic filters, and $c\geq (K+1)$ is required to keep $cN_c\geq 2H(K+1)$ mentioned in section \ref{sec2A} and $N_cf_0\geq2Hf_0$ required by Nyquist sampling theorem. In section \ref{sec2B}, we propose the SVDHPE algorithm that achieves transition band improvement in any spcific configuration of $c$ and $K$. Then, this section analyzes how the change of $c$ and $K$ influences the optimization effect, and hence recommends a configuration i.e., $c=3$, $K=2$ for the proposal after considering both optimization flexibility and response condition.

\subsection{The optimization flexibility of SVDHPE}
\label{sec3A}
\begin{figure}[tp!]
    \centering
    \includegraphics[width=0.48\textwidth]{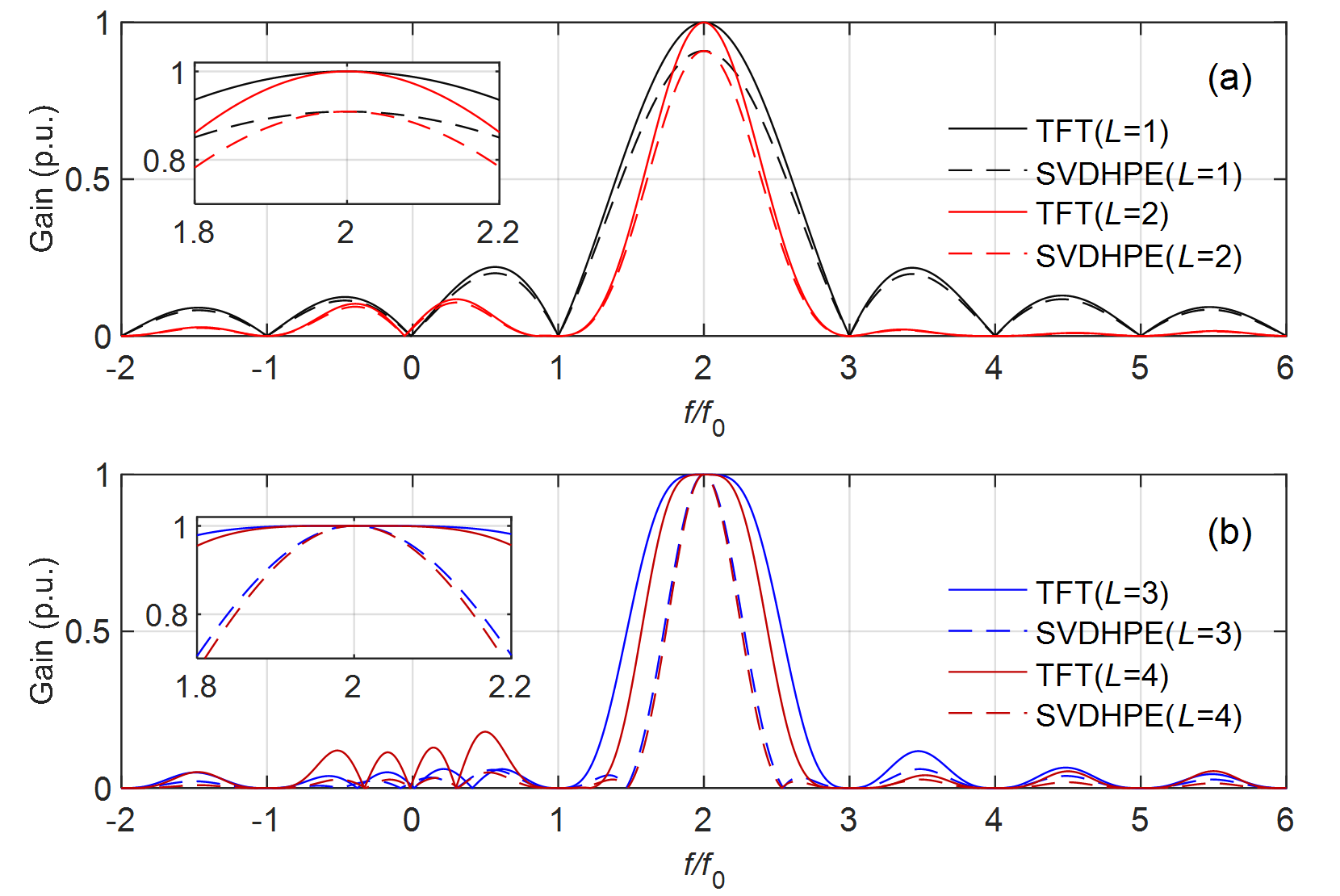}
    \caption{The gain-frequency responses of the second harmonic filter (Other harmonic filters have the same conclusion) obtained by TFT and SVDHPE algorithms when (a) $L$=1, and 2, and (b) $L$=3, and 4. The subplots are the zoom around the passband center, $f/f_0=2$.}
    \label{fig:01winlen}
\end{figure}

The optimization flexibility is characterized by the number of effective multipliers, i.e., $\lceil (K+1)/2\rceil$ where $\lceil \cdot\rceil$ is the round up operation. Besides, there is $c\geq (K+1)$. Then, compared with $c>(K+1)$, the proposal under $c=K+1$ employs less data but produces the same number of effective multipliers  when taking the same $K$. Therefore, $c=K+1=L$ is recommended in this paper. Further, when $L$ increases from one to seven referring to the window length limit in \cite{relays2018118}, the flexibility conditions are analyzed and some results are given as: 1) When $L$=1, or 2, there is only one effective multiplier $y_{h,1}$, and the dotted lines in Fig. \ref{fig:01winlen}(a) shows its influence on the harmonic phasor filter response, i.e., the passband gain being deteriorated a lot. Hence, $y_{h,1}$ is not a suitable parameter, and $L$=1, and 2 are not considered in the following analysis. 2) When $L\ge3$, there are more effective multipliers besides $y_{h,1}$, e.g., $y_{h,3}$ for $L$=3 and 4, $y_{h,3}$ and $y_{h,5}$ for $L$=5 and 6, $y_{h,3}$, $y_{h,5}$ and $y_{h,7}$ for $L$=7. When $y_{h,1}=1$, changing the rest multipliers will yield better transition band attenuation and less passband ripple deterioration as verified by Fig. \ref{fig:01winlen}(b), and the optimized multipliers are shown in Tab. \ref{tab:02}.

\subsection{The response condition of SVDHPE}
\label{sec3B}
With the multiplier values in Tab. \ref{tab:02}, we can get the optimized DHPF coefficients and test their response times. The response time is an important indicator to report the transient response ability of the phasor estimator. It means the transition time in which the total vector error (TVE) exceeds 1\% when a transient event occurs \cite{relays2018118}. In this part, we test the response times of the proposed SVDHPE under different window lengths ($c$=3, 4, 5, 6, and 7) to choose an appropriate one. The test signal is
\begin{align}
    s(t)=(1+k_{\text{a}}g(t))\cos(2\pi f_{\text{r}}t+k_{\text{p}}g(t))\nonumber\\
    +0.1(1+k_{\text{a}}g(t))\cos(2\pi hf_{\text{r}}t+k_{\text{p}}g(t)),
    \label{eq:14}
\end{align}
where the step function $g(t)=0,\,t<0$ and $g(t)=1,\,t>0$; the fundamental frequency $f_{\text{r}}=50\,$Hz; $k_{\text{a}}=0.1$, $k_{\text{p}}=0$ for amplitude step test; $k_{\text{a}}=0$, $k_{\text{p}}=-\pi/18$ for phase step test \cite{relays2018118}. For each implementation, the two components in (\ref{eq:14}) step together, and the whole test requires the harmonic order increases from 2 to $H=13$ \cite{chen2020harmonic}. 

\begin{figure}[tp!]
    \centering
    \includegraphics[width=0.48\textwidth]{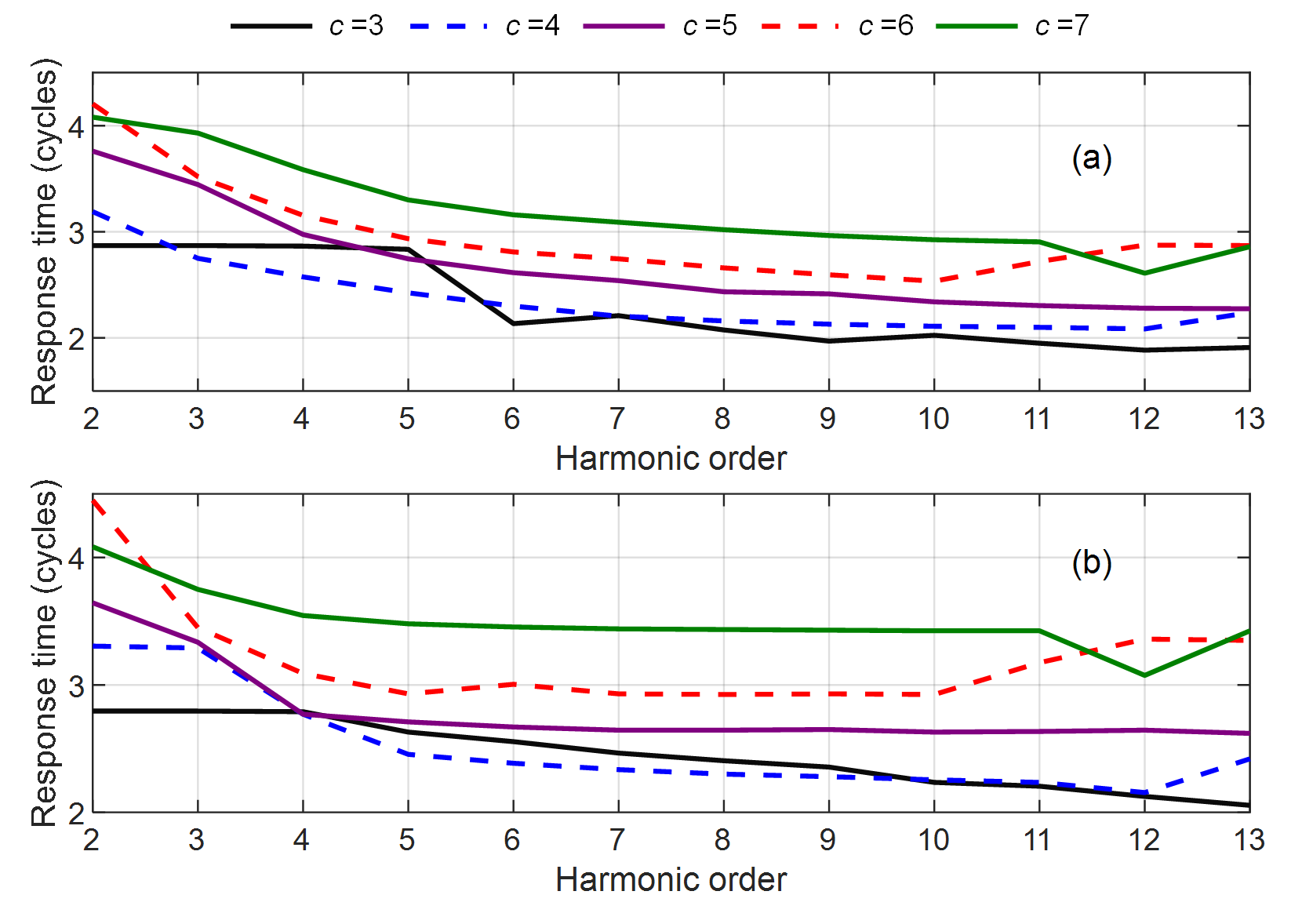}
    \caption{The response times obtained by SVDHPE algorithm under (a) amplitude step test, and (b) phase step test.}
    \label{fig:02timeres}
\end{figure}

As shown in Fig. \ref{fig:02timeres}, the results indicate that $c$=3 and $c$=4 have similar response times, and both are shorter than $c$=5, $c$=6, and $c$=7. Moreover, considering $c$=3 produces shorter response latency which is about half of the time window length \cite{relays2018118}, $c$=3 and corresponding Taylor series order $K$=2 are recommended in this paper.
\begin{table}[tp!]
    \centering
    \caption{The multiplier optimal values of SVDHPE algorithm under different $L$. Whereas $y_{h,1}$=1 is not listed.}
    \setlength{\tabcolsep}{1.5mm}
    \renewcommand\arraystretch{1.0}
    \begin{tabular}{cccccccccc}
    \toprule
	  Order &$L$=3  &$L$=4	&\multicolumn{2}{c}{$L$=5}	&\multicolumn{2}{c}{$L$=6}	&\multicolumn{3}{c}{$L$=7}\\
	  \cmidrule(lr){2-2} \cmidrule(lr){3-3} \cmidrule(lr){4-5} \cmidrule(lr){6-7} \cmidrule(lr){8-10}
	  $h$ &$y_{h,3}$ &$y_{h,3}$ &$y_{h,3}$ &$y_{h,5}$ &$y_{h,3}$ &$y_{h,5}$ &$y_{h,3}$ &$y_{h,5}$ &$y_{h,7}$ \\
	 \midrule
    2 &2.31 &1.57 &1.60 &4.47 &1.50 &3.78 &1.30 &2.40 &6.28\\
    3 &2.31 &1.59 &1.60 &4.48 &1.40 &2.96 &1.40 &3.00 &9.53	\\
    4 &2.31 &1.60 &1.50 &3.71 &1.40 &2.98 &1.40 &3.00 &9.53	\\
    5 &2.31 &1.60 &1.50 &3.71 &1.40 &2.98 &1.40 &3.00 &9.52	\\
    6 &2.31 &1.60 &1.50 &3.71 &1.40 &2.98 &1.40 &3.00 &9.53	\\
    7 &2.31 &1.60 &1.50 &3.71 &1.40 &2.98 &1.40 &3.00 &9.53	\\
    8 &2.31 &1.60 &1.50 &3.71 &1.40 &2.97 &1.40 &3.00 &9.53	\\
    9 &2.31 &1.59 &1.50 &3.71 &1.40 &2.96 &1.40 &3.00 &9.54	\\
    10 &2.31 &1.58 &1.50 &3.70  &1.40 &2.94 &1.40 &3.00 &9.54	\\
    11 &2.32 &1.55 &1.50 &3.70  &1.50 &3.75 &1.40 &3.00 &9.55	\\
    12 &2.32 &1.47 &1.50 &3.69  &1.60 &4.94 &1.30 &2.40 &6.26	\\
    13 &2.24 &1.81 &1.50 &3.68  &1.60 &5.11 &1.40 &3.02 &9.52	\\
    \bottomrule
    \end{tabular}
    \label{tab:02}
\end{table}

\begin{figure}[tp!]
    \centering
    \includegraphics[width=0.48\textwidth]{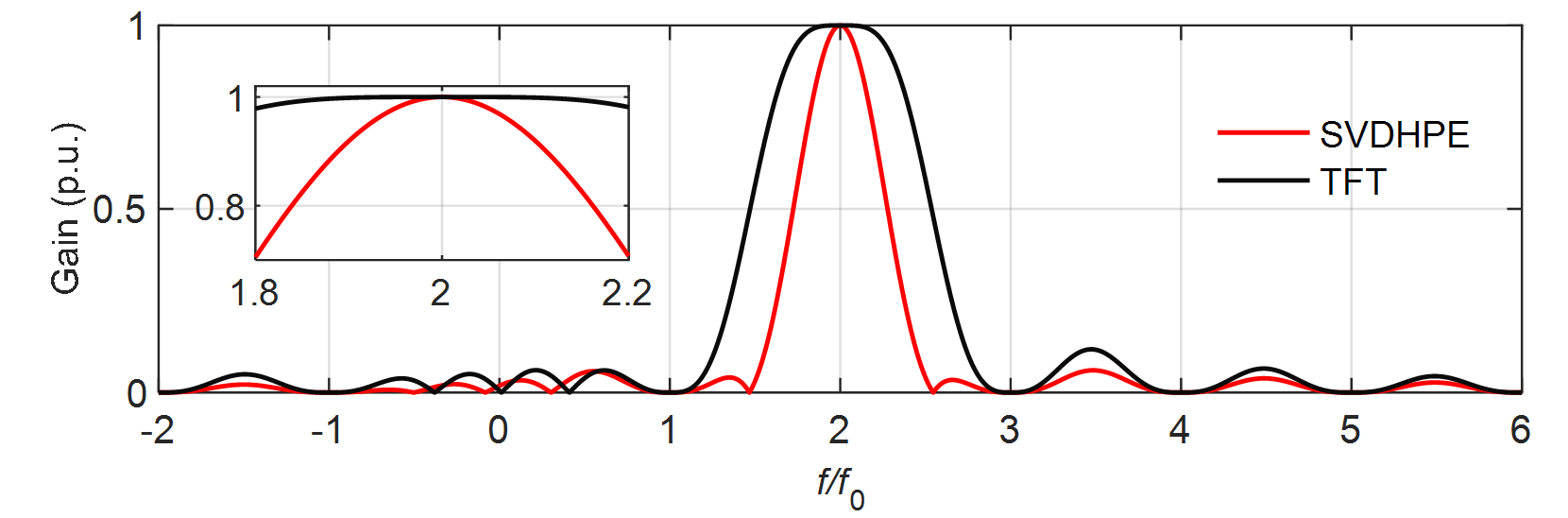}
    \caption{The gain-frequency response of the second harmonic filter obtained by TFT, and SVDHPE algorithms when $c$=3, $K$=2.}
    \label{fig:03filter}
\end{figure}

With the suggested $c$=3 and $K$=2, we analyze the maximum transition band gain for all 2-13 \cite{chen2020harmonic} order harmonics obtained by the TFT and SVDHPE algorithms, and the results are shown in Tab. \ref{tab:03}. Compared with TFT, the proposed SVDHPE reduces the maximum transition band gain by more than 86\%. The gain-frequency response in Fig. \ref{fig:03filter} intuitively verifies this improvement. In addition, Fig. \ref{fig:03filter} shows the proposed SVDHPE does not keep the whole flat gain of the TFT. Other harmonic filters have a similar trend, and the effects will be shown in the following test \ref{sec4D}.

\begin{table}[tp!]
    \centering
    \caption{The maximum transition band gain and reduction for all 2-13 order harmonics obtained by TFT and SVDHPE algorithms when $c$=3, $K$=2.}
    \setlength{\tabcolsep}{1.8mm}
    \renewcommand\arraystretch{1.0}
    \begin{tabular}{cccc}
    \toprule
	Order $h$  &TFT  &SVDHPE	&Reduction	\\
	  \midrule
    2 &0.5800 &0.0410 &92.9\%	\\
    3 &0.5771 &0.0384 &93.3\%	\\
    4 &0.5764 &0.0377 &93.5\%	\\
    5 &0.5761 &0.0374 &93.5\%	\\
    6 &0.5761 &0.0375 &93.5\%	\\
    7 &0.5762 &0.0378 &93.4\%	\\
    8 &0.5762 &0.0380 &93.4\%	\\
    9 &0.5763 &0.0384 &93.3\%	\\
    10 &0.5766 &0.0394 &93.2\% \\
    11 &0.5771 &0.0412 &92.9\% \\
    12 &0.5787 &0.0452 &92.2\% \\
    13 &0.5824 &0.0802 &86.2\% \\
    \bottomrule
    \end{tabular}
    \label{tab:03}
\end{table}

\section{Numerical Tests}
\label{sec4}
Referring to the IEC/IEEE standard \cite{relays2018118}, some typical scenarios are simulated to test the OBI suppression ability of the proposed SVDHPE algorithm. The MP and ESPRIT algorithms are compared for having high frequency resolution and good attenuation to OBI tones, while the TFT is also compared to show the improvement of SVDHPE. For a fair comparison, the four algorithms have the same test settings including: the sampling frequency $f_\text{s}=10\,$kHz, the reporting rate $f_{\text{re}}=50\,$frames/sec, the nominal frequency $f_0=50\,$Hz, and the time window $T_{\text{w}}=3/f_0$. Besides, the Taylor order $K=2$ for SVDHPE and TFT \cite{platas2010dynamic}; the threshold $\rho=10^{-3}$ for MP to distinguish the signal and noise singular values \cite{sheshyekani2016general}. Besides, there will be some outliers in the maximum estimation errors of the MP and ESPRIT algorithms when the signal component frequencies are not well identified. For readability, these outliers are discarded.

\subsection{Different interharmonic amplitude test}
\label{sec4A}
This test employs the input signal with the form of
\begin{align}	
	\resizebox{0.95\hsize}{!}{$
    s(t)=\sum_{h=1}^{H}A_h\cos (2\pi hft+\phi_h)+\sum_{h=2}^{H}A_{h,i}\cos (2\pi f_{h,i}t+\phi_{h,i}),$}
    \label{eq:15}
\end{align}
where $A_1,\,f,\,\phi_1$ are respectively the fundamental amplitude, frequency, and phase; $A_h,\,hf,\,\phi_h\,(h\in[2,\,H])$ represent the $h-$order harmonic amplitude, frequency, and phase, where the harmonic frequency is set to an integer multiple of the fundamental, and the maximum order $H=13$ \cite{chen2020harmonic}; $A_{h,i},\,f_{h,i},\,\phi_{h,i}$ denote the amplitude, frequency, and phase of OBI  tone close to $h-$order harmonic.

The signal in (\ref{eq:15}) contains fundamental, 12 harmonics (from second to thirteenth), and 12 OBI tones. The signal settings are: $f=f_0$; the OBI frequencies $f_{h,i}=hf_0-0.5f_{\text{re}}$; the harmonic amplitudes are 10\% of the fundamental, i.e., $A_1=1\,\text{p.u.},\,A_h=0.1\,\text{p.u.}$; all the OBI components have the amplitudes increasing from 1\% to 50\% of the harmonics, i.e., $A_{h,i}\in[0.001,\,0.05]\,\text{p.u.}$; the phases of the fundamental, harmonic, and OBI, $\phi_1,\,\phi_h,\,\phi_{h,i}$ are randomly chosen in $[-\pi,\,\pi]$. Note in the subsequent tests, the settings about signal phases, fundamental amplitude, and the OBI frequencies are the same as here, and they will not be repeatedly stated.

\begin{figure}[tp!]
    \centering
    \includegraphics[width=0.48\textwidth]{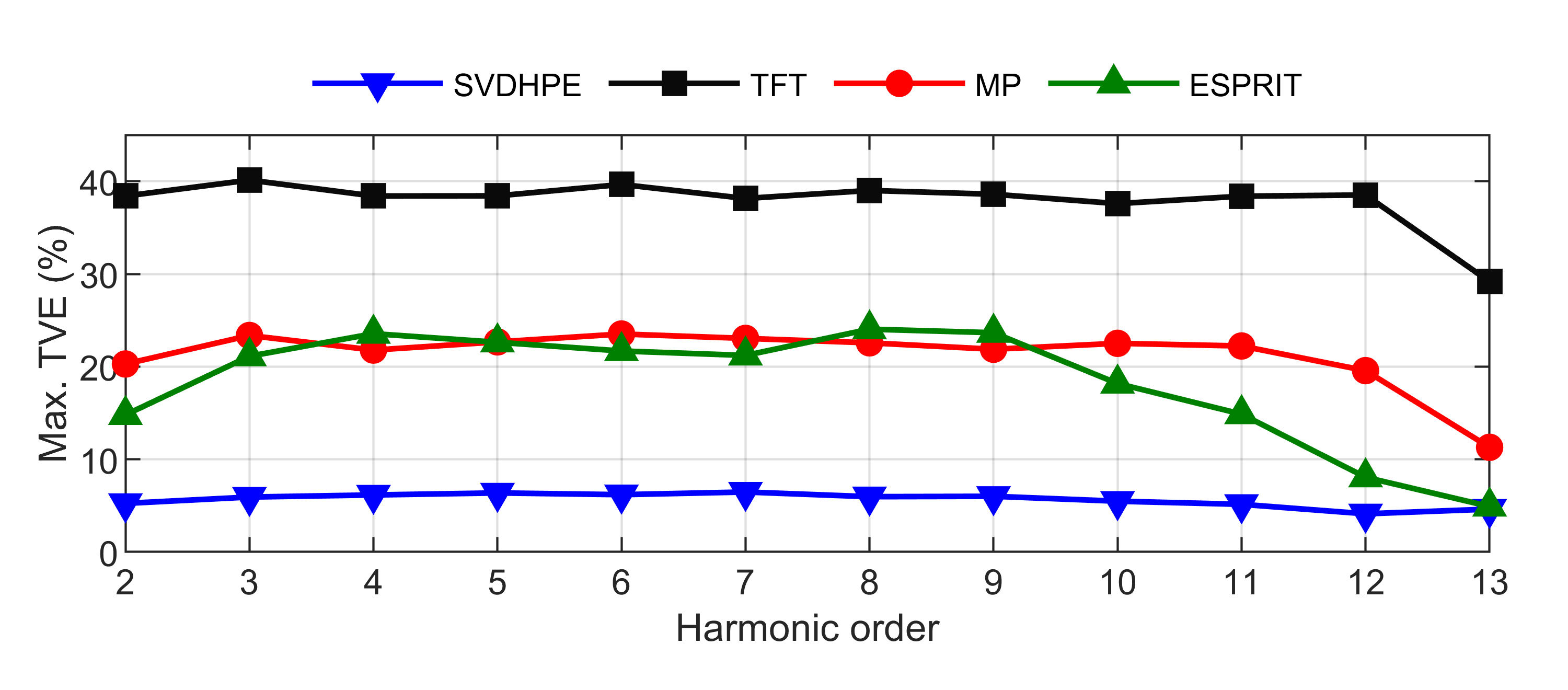}
    \caption{Maximum harmonic TVE obtained by SVDHPE, TFT, MP, and ESPRIT algorithms under different OBI amplitude.}
    \label{fig:04obiamp}
\end{figure}

Then, we employ the proposed SVDHPE, TFT, MP, and ESPRIT algorithms to extract the 2-13 harmonic phasors while the OBI amplitudes increase by 0.001\,p.u. in each test. For the whole OBI amplitude range, i.e., 0.001 to 0.05\,p.u., the maximum TVEs of all harmonic phasors obtained by the four algorithms are shown in Fig. \ref{fig:04obiamp}. The proposed SVDHPE achieves the best estimation among the four algorithms. Even all the OBI amplitudes come up to 50\% of the harmonics, its TVEs for all harmonic phasors are always well below 6.5\%, and are over 83\% less than the TFT algorithm.

\subsection{Different harmonic amplitude test}
This test applies the same signal model as shown in (\ref{eq:15}). Namely, the fundamental, 2-13 harmonics and some OBI tones are set as the input of the four algorithms. The fundamental and harmonic frequencies $hf=hf_0$; the OBI amplitudes $A_{h,i}=0.01\,\text{p.u.}$; the harmonic amplitudes increase from 8\% to 12\% of the fundamental with a step of 0.5\% in each test, i.e., $A_{h}\in[0.08,\,0.12]\,\text{p.u.}$.

\begin{figure}[tp!]
    \centering
    \includegraphics[width=0.48\textwidth]{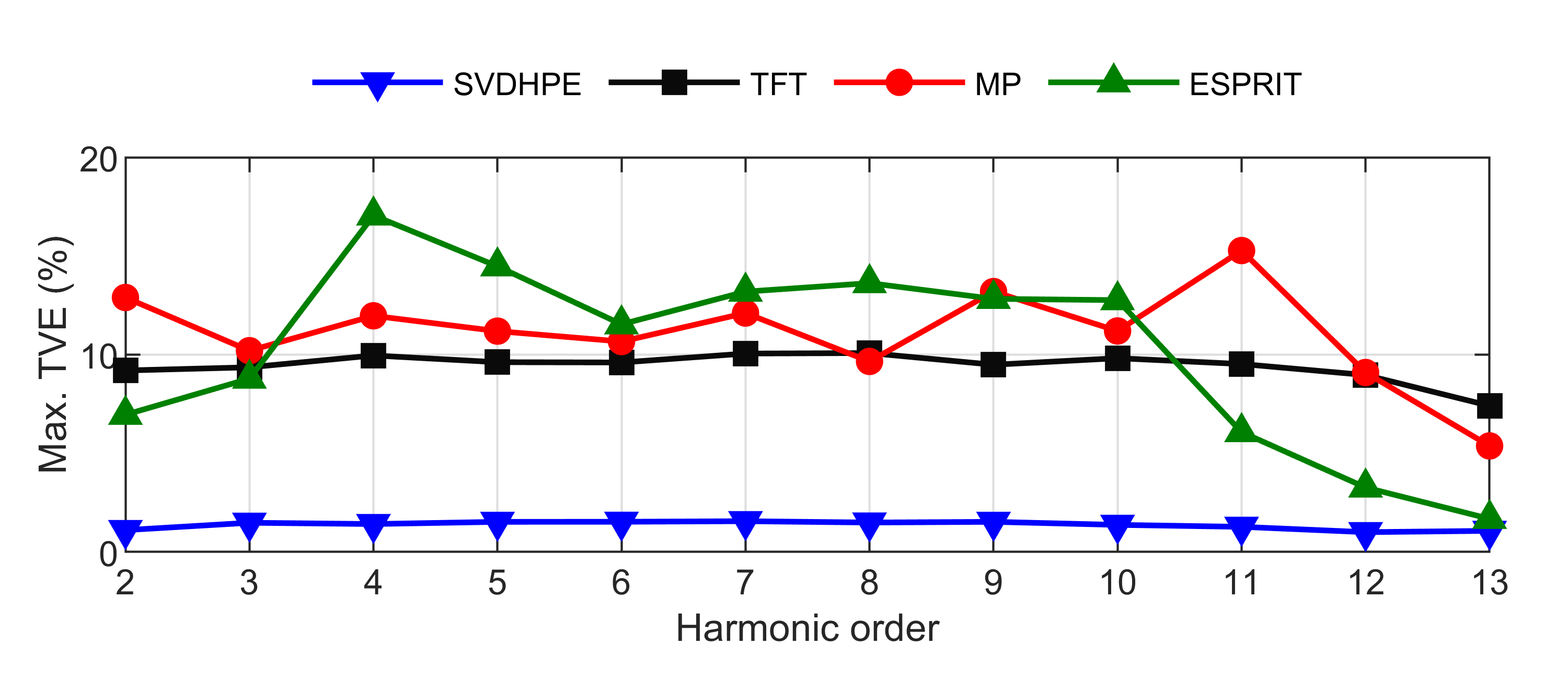}
    \caption{Maximum harmonic TVE obtained by SVDHPE, TFT, MP, and ESPRIT algorithms under different harmonic amplitude.}
    \label{fig:05haramp}
\end{figure}

Fig. \ref{fig:05haramp} shows the maximum TVEs for 2-13 harmonic phasors obtained by SVDHPE, TFT, MP, and ESPRIT algorithms over the whole harmonic amplitude range. The proposed SVDHPE achieves TVEs well below 1.6\%, indicating it is robust against harmonic amplitude changes.

\subsection{Noise interference and OBI test}
The signal in (\ref{eq:15}) added with a white noise is used in this test. The signal-to-noise ratio (SNR) is defined as
\begin{align}
    \text{SNR}=10\log\frac{A_1^2}{2\sigma^2},
    \label{eq:16}
\end{align}
where $A_1=1\,$p.u. is the fundamental amplitude in (\ref{eq:15}); $\sigma$ is the variance of the white noise.

\begin{figure}
    \centering
    \includegraphics[width=0.48\textwidth]{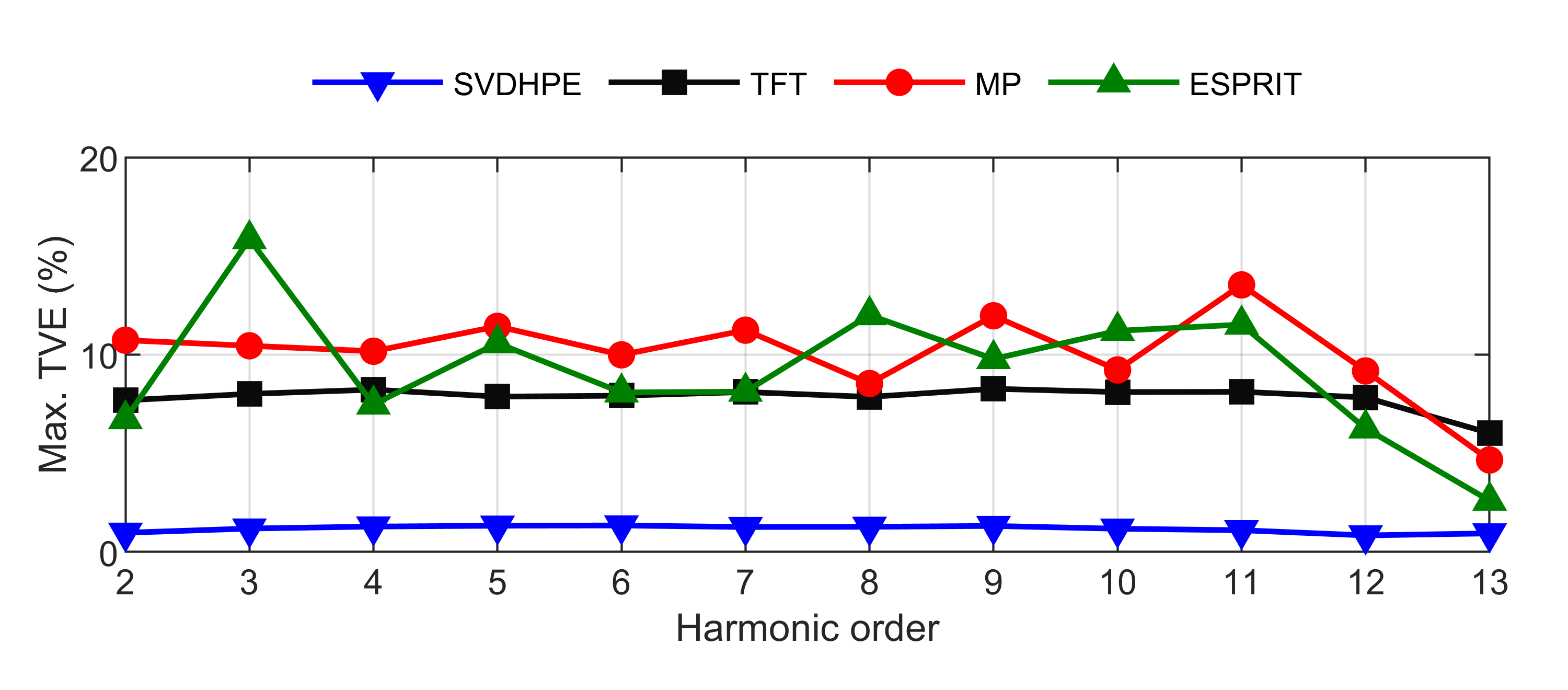}
    \caption{Maximum harmonic TVE obtained by SVDHPE, TFT, MP, and ESPRIT algorithms under noise and OBI conditions.}
    \label{fig:06noise}
\end{figure}

The signal settings include: the frequencies of the fundamental and harmonics are $
hf_0\,$; the OBI amplitudes are 10\% of the harmonics, and the harmonic amplitudes are 10\% of the fundamental, i.e., $A_{h,i}=0.01\,\text{p.u.},\,A_{h}=0.1\,\text{p.u.}$; the SNR defined in (\ref{eq:16}) is set from 50\,dB to 80\,dB with a step of 5\,dB in each test. As seen in Fig. \ref{fig:06noise}, the proposed SVDHPE shows good noise immunity for the maximum TVE values being less than 1.4\%, and again holds the best performance among these four algorithms.

\subsection{Harmonic frequency deviation and OBI test}
\label{sec4D}
The test signal contains the fundamental, 2-13 harmonics, and some OBI tones, as the form of (\ref{eq:15}). The fundamental frequency $f$ increases from 49.5\,Hz to 50.5\,Hz with a step of 0.1\,Hz for each test. As a result, the harmonic frequency changes within $[49.5h,\,50.5h]$ at a step of $0.1h$. The harmonic amplitude $A_h=0.1\,\text{p.u.}$, the OBI amplitude $A_{h,i}=0.01\,\text{p.u.}$.

\begin{figure}[tp!]
    \centering
    \includegraphics[width=0.48\textwidth]{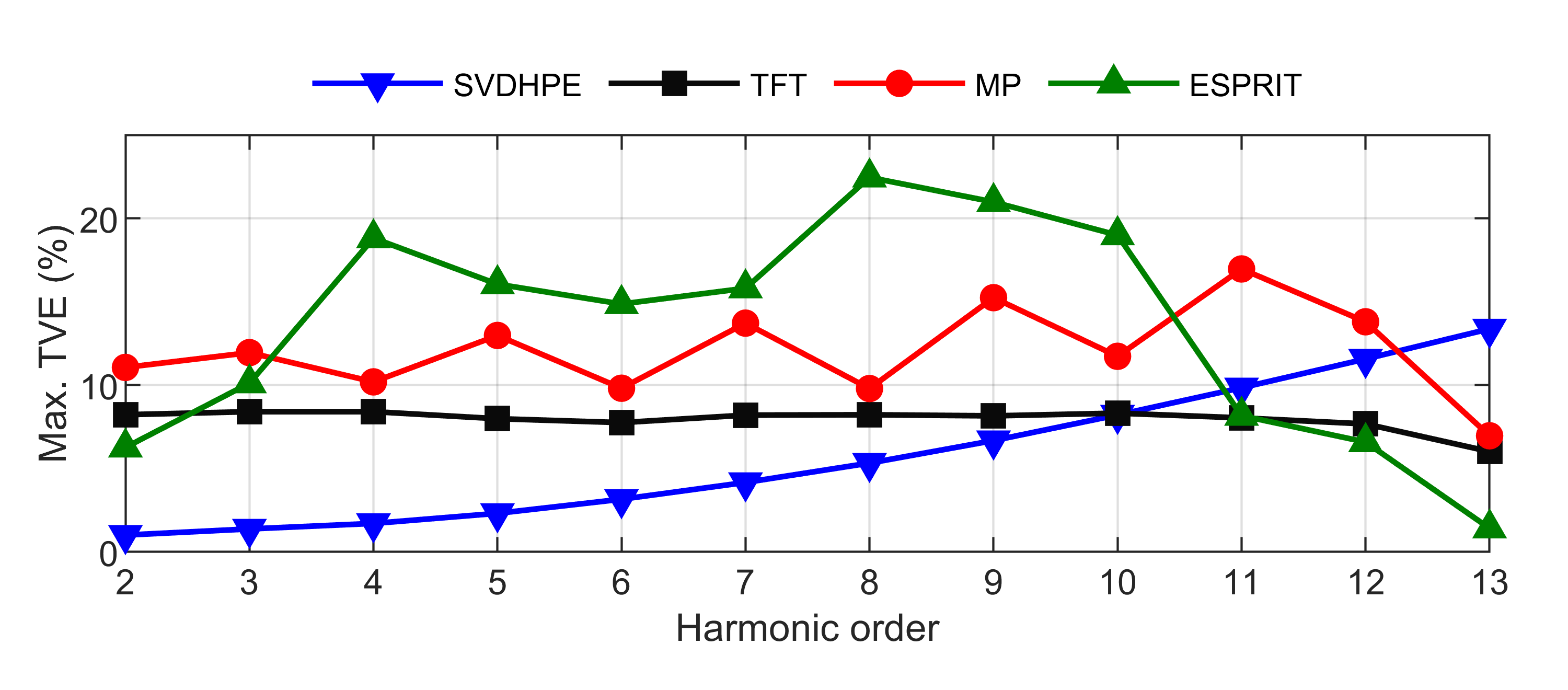}
    \caption{Maximum harmonic TVE obtained by SVDHPE, TFT, MP, and ESPRIT algorithms under harmonic frequency deviation and OBI conditions.}
    \label{fig:07fredev}
\end{figure}

Fig. \ref{fig:07fredev} shows the maximum estimation errors throughout the frequency deviation range for 2-13 harmonics obtained by the four algorithms. In this case, the proposed SVDHPE has the maximum TVEs increasing with the harmonic order, and hence greater TVEs than TFT when harmonic order exceeds ten. This is because the proposed SVDHPE results in losing harmonic filter flat gain as seen in Fig. \ref{fig:03filter}. However, the SVDHPE still has advantages over the other three algorithms when the harmonic order is no more than ten. Therefore, the proposal can be used to identify the distribution topology \cite{chen2020switch}, high impedance fault \cite{farajollahi2017location}, and so on where the third, fifth, and seventh harmonic phasors are employed.

\subsection{Harmonic amplitude modulation and OBI test}
The following three simulations are carried out to test the dynamic performance of proposed SVDHPE. For the harmonic amplitude modulation and OBI test, the signal contains the fundamental, all harmonics from 2 to 13, and the corresponding interharmonics, i.e.,
\begin{align}
    s(t)=&\sum_{h=1}^{H}A_h(1+k_{\text{m,a}}\cos(2\pi f_\text{m}t))\cos (2\pi hft+\phi_h)\nonumber\\
    &+\sum_{h=2}^{H}A_{h,i}\cos (2\pi f_{h,i}t+\phi_{h,i}).
    \label{eq:17}
\end{align}

\begin{figure}[tp!]
    \centering
    \includegraphics[width=0.48\textwidth]{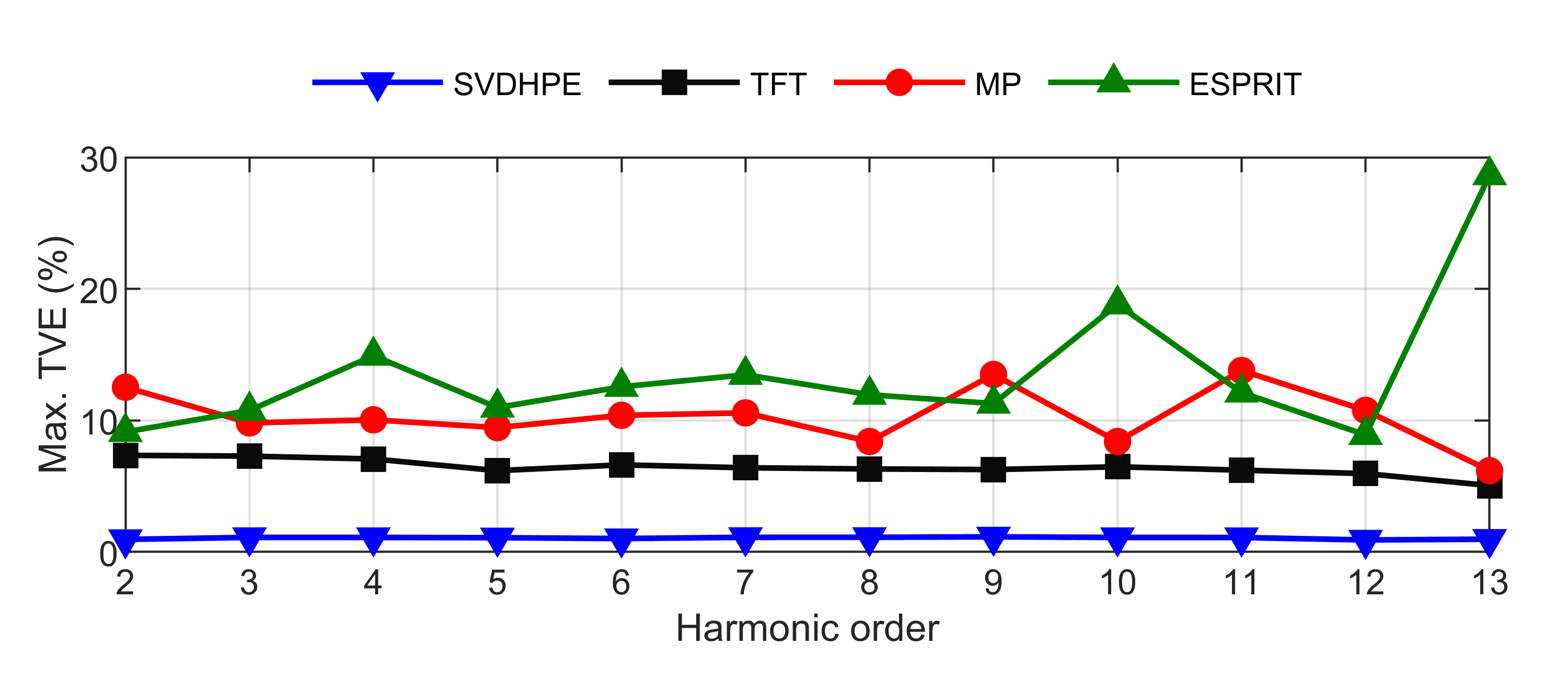}
    \caption{Maximum harmonic TVE obtained by SVDHPE, TFT, MP, and ESPRIT algorithms under harmonic amplitude modulation and OBI conditions.}
    \label{fig:08haram}
\end{figure}

The frequencies of the fundamental and harmonic are $hf_0$. The amplitudes of the fundamental, harmonic, and OBI are $A_1=1\,\text{p.u.}$, $A_h=0.1\,\text{p.u.}$, and $A_{h,i}=0.01\,\text{p.u.}$, respectively. The fundamental and harmonic amplitudes are modulated with level $k_{\text{m,a}}=0.1h\,(h\in[1,\,H])$, and frequency $f_\text{m}\in[0.1,\,2]$\,Hz at a change of 0.1\,Hz for every test. Fig. \ref{fig:08haram} shows the maximum TVEs of 2-13 harmonics obtained by the SVDHPE, TFT, MP, and ESPRIT algorithms over the whole modulating range. It is observed that the proposal has good dynamic performance for realizing estimation errors always below 1.2\%.

\subsection{Harmonic phase modulation and OBI test}
\begin{figure}[tp!]
    \centering
    \includegraphics[width=0.48\textwidth]{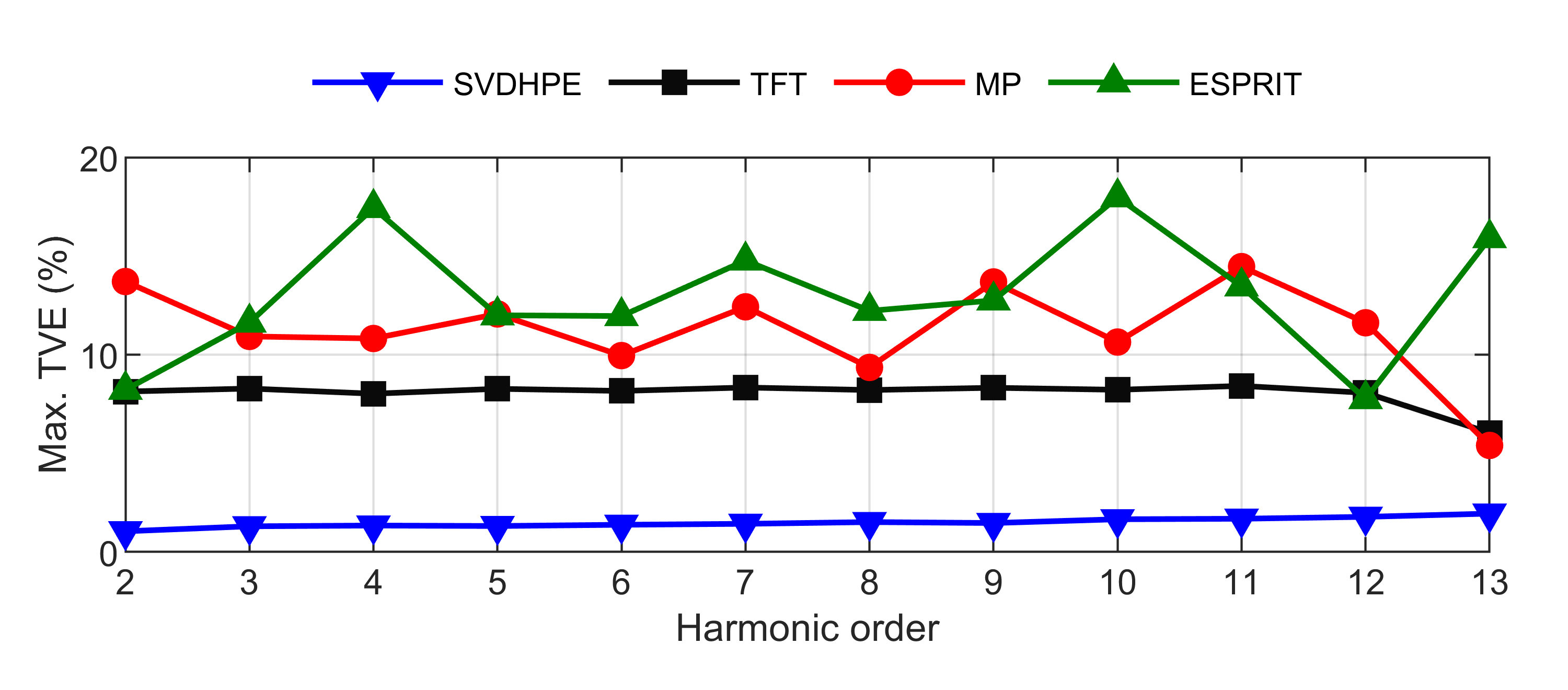}
    \caption{Maximum harmonic TVE obtained by SVDHPE, TFT, MP, and ESPRIT algorithms under harmonic phase modulation and OBI conditions.}
    \label{fig:09fhpm}
\end{figure}

In this test, the fundamental and harmonic (from second to thirteenth) phases are added with modulation components, i.e.,
\begin{align}
    s(t)=&\sum_{h=1}^{H}A_h\cos (2\pi hft+k_{\text{m,p}}\cos(2\pi f_\text{m}t-\pi)+\phi_h)\nonumber\\
    &+\sum_{h=2}^{H}A_{h,i}\cos (2\pi f_{h,i}t+\phi_{h,i}),
    \label{eq:18}
\end{align}
where the fundamental frequency $f=f_0$; the fundamental, harmonic, and OBI have amplitudes $A_1=1\,\text{p.u.}$, $A_h=0.1\,\text{p.u.}$, and $A_{h,i}=0.01\,\text{p.u.}$, respectively. For each test, the phase modulation level $k_{\text{m,p}}=0.1h\,(h\in[1,\,H])$, and the modulation frequency increases from 0.1\,Hz to 2\,Hz with a step of 0.1\,Hz. Namely, the real-time harmonic frequencies oscillate at a form of $hf_0-0.1hf_\text{m}\sin(2\pi f_\text{m}t-\pi)$ in each test. Again, Fig. \ref{fig:09fhpm} shows the proposed SVDHPE achieves the smallest estimation errors for all 2-13 harmonic phasors (maximum TVEs below 2\%).

\subsection{Harmonic frequency ramp and OBI test}
As (\ref{eq:19}) shows, the test signal includes the fundamental, 2-13 harmonics, and the related OBIs. For each test, the fundamental frequency $f+R_\text{f}t$ increases from 49.5\,Hz to 50.5\,Hz in one second with the frequency ramp $R_\text{f}=1\,$Hz/s. As a result, the harmonic frequency changes within $[49.5h,\,50.5h]$ with the frequency ramp $hR_\text{f}=h\,$Hz/s\,($h\in[2,\,H]$). The harmonic amplitude $A_h=0.1\,\text{p.u.}$, and the OBI amplitude $A_{h,i}=0.01\,\text{p.u.}$.
\begin{align}
    s(t)=&\sum_{h=1}^{H}A_h\cos (2\pi hft+\pi hR_\text{f}t^2+\phi_h)\nonumber\\
    &+\sum_{h=2}^{H}A_{h,i}\cos (2\pi f_{h,i}t+\phi_{h,i}).
    \label{eq:19}
\end{align}

\begin{figure}[tp!]
    \centering
    \includegraphics[width=0.48\textwidth]{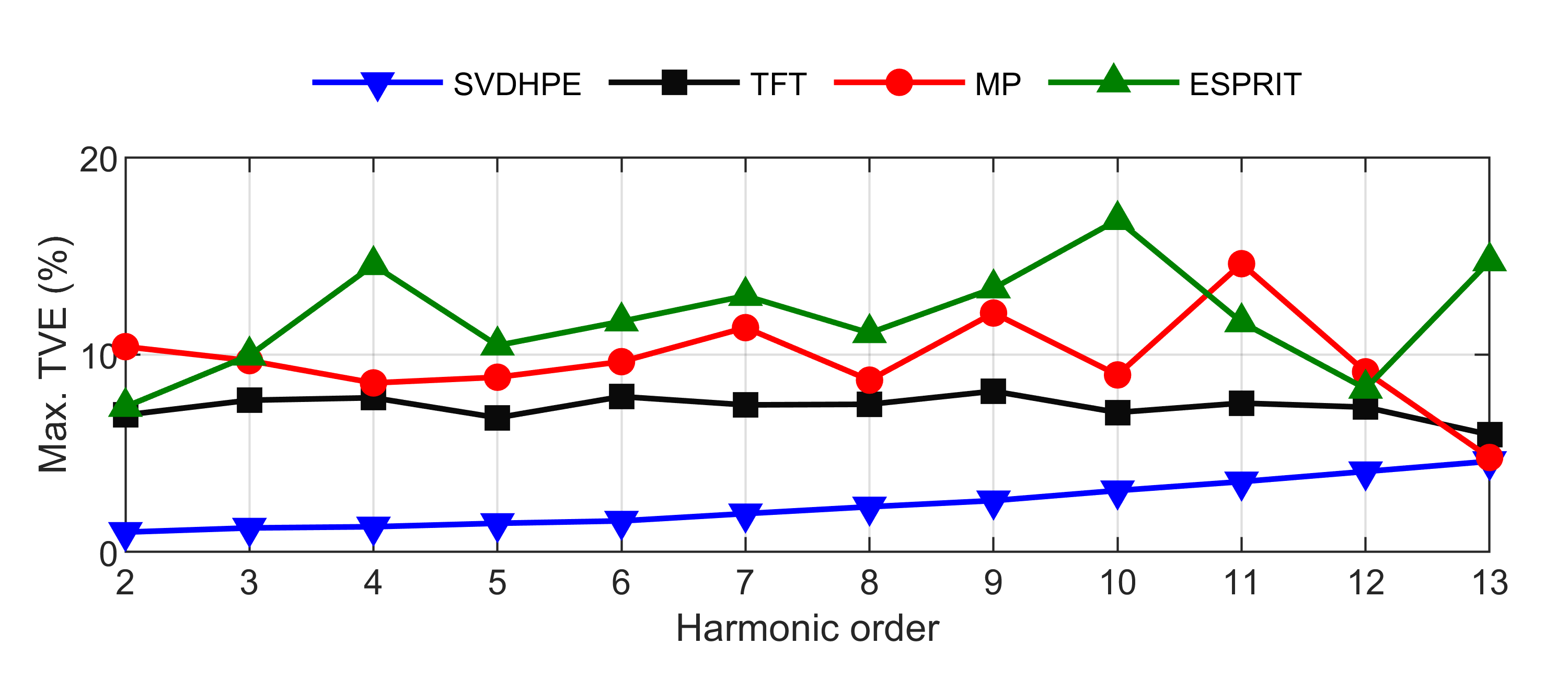}
    \caption{Maximum harmonic TVE obtained by SVDHPE, TFT, MP, and ESPRIT algorithms under harmonic frequency ramp and OBI conditions.}
    \label{fig:10ramp}
\end{figure}

Fig. \ref{fig:10ramp} shows similar results to the case \ref{sec4D} harmonic frequency deviation and OBI test, i.e. the maximum estimation errors of the proposed SVDHPE increase with the harmonic order for the increasing passband ripple. In this test, however, the proposed SVDHPE still performs the best for all 2-13 harmonics because the frequency offsets at most of the time are less than in test \ref{sec4D}. Therefore, the negative effects caused by the deteriorated passband performance are limited.

As seen in the tests A-G, the proposed harmonic phasor estimator produces maximum TVEs higher than 1\% required by the synchrophasor standard \cite{relays2018118}. Because the test signals contain the changing harmonic parameter, and the interference from fundamental leakage, other 11 harmonics, and 12 OBI tones at the same time, which is more complex than the signal in the standard \cite{relays2018118}. Moreover, the proposed SVDHPE performs better estimation accuracy than the TFT, MP and ESPRIT algorithms in most conditions, which shows the proposal is an useful exploration for harmonic phasor estimation enabled with suppression of interharmonics. Besides, the proposal achieves maximum TVEs below 5\% for all 2-8 harmonic phasors under all tests A-G. This accuracy can meet the requirement of practical applications, e.g. the identification of distribution topology \cite{chen2020switch}, and high impedance fault \cite{farajollahi2017location} where the third, fifth, and seventh harmonic phasors are most commonly used.

\section{Experimental Test}
\label{sec5}
\begin{figure}[tp!]
	\centering
	\includegraphics[width=0.48\textwidth]{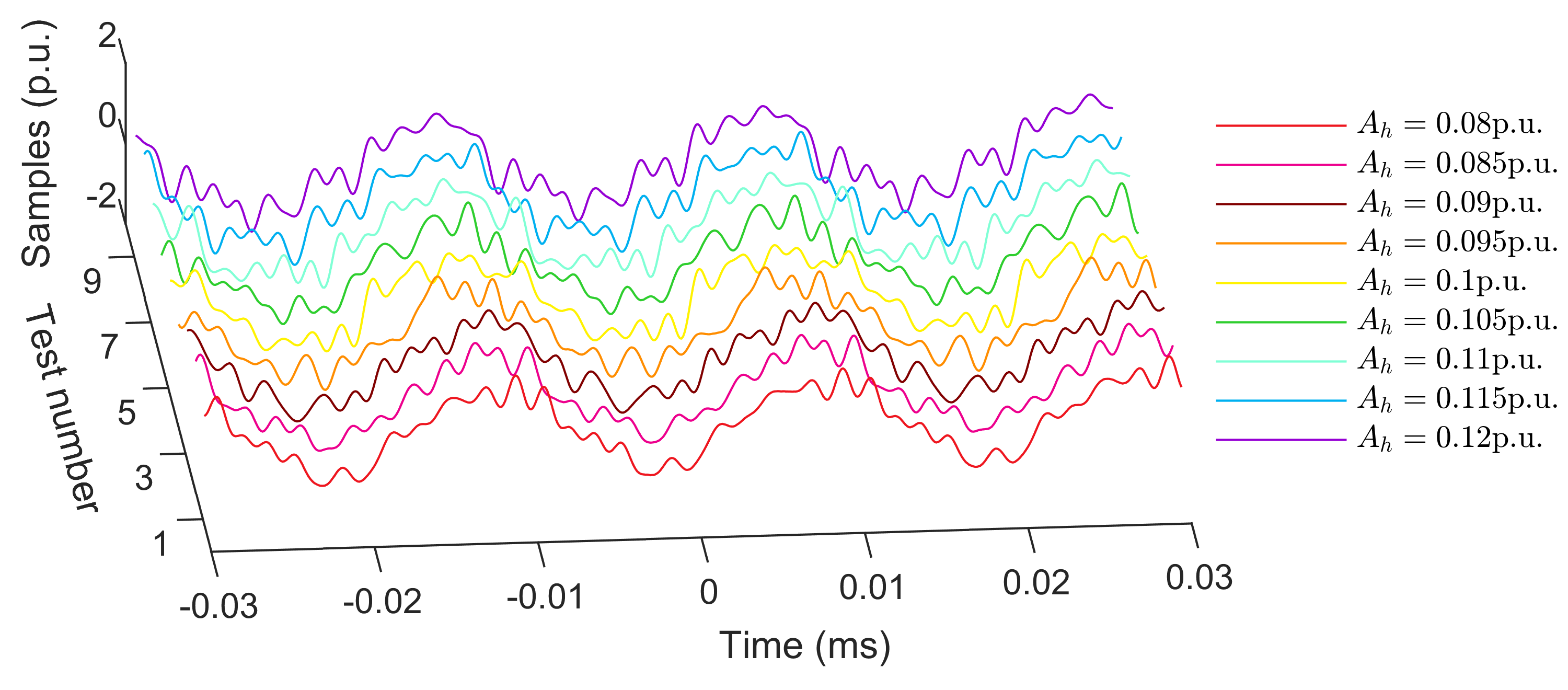}
	\caption{The samples of 9 test signals in the experimental test.}
	\label{fig:13expdata}
\end{figure}

The platform is composed of a signal generator (Tektronix AFG 31252), and a digital voltmeter (Keysight 3458A). The experimental test employs the signal settings in case B because the harmonic amplitude change is common in practice. In each test, we use the signal generator to produce all the 25 signal components in (\ref{eq:15}) one by one, and use the voltmeter to get their samples. Then, the signal samples of the 25 components are added together as shown in (\ref{eq:15}). When the harmonic amplitudes increase from 0.08\,p.u. to 0.12\,p.u. in a step of 0.005\,p.u., there are 9 test singals, and their samples of one time window are shown in Fig. \ref{fig:13expdata}. Finally, these distorted signals are processed by the SVDHPE, TFT, MP, and ESPRIT algorithms to obtain the harmonic phasors. To characterize the estimation accuracy, a new index, the signal residual in a time window is applied, i.e.,
\begin{align}
	\text{Res}=\sqrt{\frac{\sum_{n=-N_h}^{N_h}(s_h(t+nT_{\text{s}})-\hat{s}_h(t+nT_{\text{s}}))^2}{\sum_{n=-N_h}^{N_h}s_h^2(t+nT_{\text{s}})}},
	\label{eq:20}
\end{align}
where $s_h(t+nT_{\text{s}})$ denotes the samples of $h-$order harmonic; $\hat{s}_h(t+nT_{\text{s}})$ means the constructed samples using the estimated harmonic phasors obtained by different estimators.

\begin{figure}[tp!]
	\centering
	\includegraphics[width=0.48\textwidth]{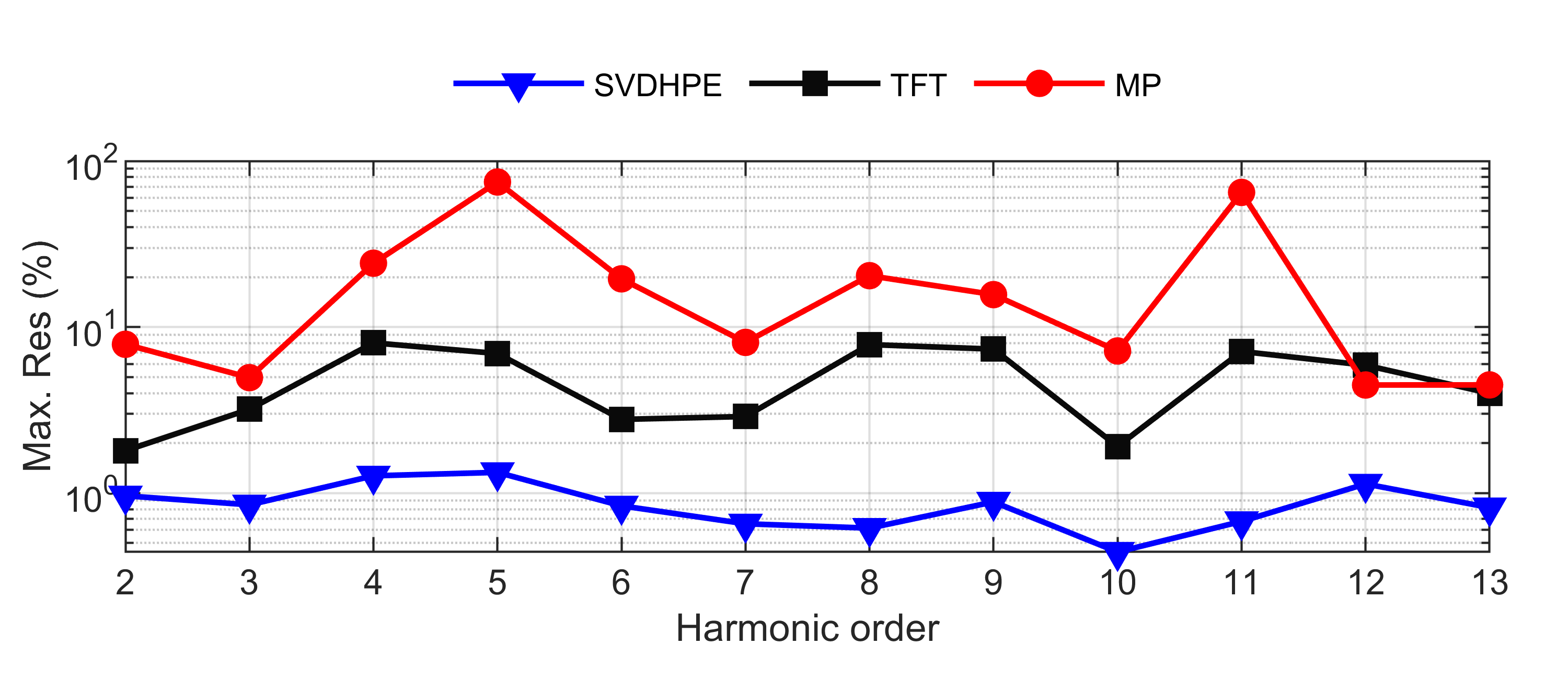}
	\caption{Maximum harmonic Res obtained by SVDHPE, TFT, and MP algorithms under the experimental test. The ESPRIT algorithm can not identify any harmonic tone in this test, and hence there is no curve to show for this method.}
	\label{fig:13exp}
\end{figure}

For all 12 harmonic phasors, the maximum residuals obtained by the four algorithms are shown in Fig. \ref{fig:13exp}. There is no curve for ESPRIT algorithm because it can not identify any harmonic tone in this test. As shown in Fig. \ref{fig:13exp}, the proposed SVDHPE algorithm still keeps the best estimation accuracy under the four algorithms. Besides, the proposal yields the maximum residual below 2.1\% which is close to the maximum TVE 1.6\% in case B, proving the effectiveness of the SVDHPE algorithm.

\section{Analysis of the Computational Burden}
Tab. \ref{tab:04} shows the computational burden analysis for the four algorithms compared in the sections \ref{sec4} and \ref{sec5}. The parameters $N,\,M,\,H$ represent the sample number in a time window, the number of all signal components, and the maximum harmonic order, respectively. The main real-time computation for the TFT and SVDHPE algorithms comes from the (\ref{eq:11}). Compared with TFT algorithm, the proposed SVDHPE does not increase online computation, and hence keep the efficient computation. But for the MP and ESPRIT algorithms, the real-time filter  design according to the solved component frequencies, i.e., the online calculation of (\ref{eq:5}) increases much time complexity \cite{song2021accurate}. More intuitively, we carry out a test 10000 running by Matlab R2020a on a computer with a 16 GB RAM and a 3.2 GHz processor. The test signal contains the fundamental, 2-13 harmonics, and the related OBI tones as in case \ref{sec4A}. The average execution times for one data frame of the four algorithms are also shown in Tab. \ref{tab:04}. It can be seen that the proposed SVDHPE needs very slight real-time execution time, which helps to serve the real-time applications.

\begin{table}[tp!]
    \centering
    \caption{The computation analysis of the proposed SVDHPE, TFT, MP, and ESPRIT algorithms.}
    \setlength{\tabcolsep}{1.8mm}
    \renewcommand\arraystretch{1.0}
    \begin{tabular}{ccc}
    \toprule
	Algorithms  &Time complexity  &Execution time (ms)	\\
	  \midrule
    SVDHPE   &$O(2N(H-1))$          &0.22	\\
    TFT      &$O(2N(H-1))$          &0.25	\\
    MP       &$O(4M(2N+1)^2)$   &82.42	\\
    ESPRIT   &$O(4M(2N+1)^2)$   &14.27	\\
    \bottomrule
    \end{tabular}
    \label{tab:04}
\end{table}

\section{Conclusion}
With the increasing signal distortion caused by the diffusion of power electronic devices and nonlinear loads, it is necessary to develop accurate dynamic harmonic phasor estimator immune from interharmonics to better serve the related applications. This paper proposes the decomposition form for harmonic filters by applying the singular value decomposition to the Taylor-Fourier transform algorithm, and constructs an optimization problem to design dynamic harmonic phasor filters with high suppression of out-of-band interference. A set of tests are carried out under input signals that contain multiple harmonics, and interharmonic tones accompanied by noise, harmonic frequency deviation, amplitude modulation, phase modulation, and frequency ramp. The results verify that the proposed algorithm indeed achieves excellent harmonic phasor estimation under these tests, because the introduced variables have considerably improved the transition band attenuation of the harmonic phasor filters. Besides, this paper considers the online computation, and reporting rate to make the proposed algorithm applicable to hardware devices. In the future work, the specifications dedicated for harmonic phasors, and the influences of hardware devices’ sampling rate, memory and computation capability need to be further explored.

\section*{Appendix}
This section proves that the odd elements in the first row of right singular matrix $\boldsymbol{D}$ are not zero and the even ones equal zero, i.e., $d_{1(2a+1)}\neq 0\enskip(0\le a\le\lfloor K/2\rfloor)$ and $d_{1(2a)}=0\enskip(1\le a\le\lfloor(K+1)/2\rfloor)$ where $\lfloor\cdot\rfloor $ denotes the round down operation; $K$ represents the Taylor series order.

There are two key points to prove this finding. 1) After substituting the SVD in (\ref{eq:6}) into the matrix $\boldsymbol{B}_K^\text{T}\boldsymbol{B}_K$, (\ref{eq:A1}) shows that the eigenvector matrix of $\boldsymbol{B}_K^\text{T}\boldsymbol{B}_K$ is exactly the right singular matrix of $\boldsymbol{B}_K$ \cite{kalman2018svd}. Then, the above finding is also about the first row elements in the eigenvector matrix of $\boldsymbol{B}_K^\text{T}\boldsymbol{B}_K$.
\begin{align}
    \boldsymbol{B}_K^\text{T}\boldsymbol{B}_K=\boldsymbol{D\it{\Lambda}}^{\text{T}}\boldsymbol{C}^\text{T}\boldsymbol{C\it{\Lambda D}}^\text{T}=\boldsymbol{D\it{\Lambda}}^{\text{T}}\boldsymbol{\it{\Lambda D}}^\text{T}.
    \label{eq:A1}
\end{align}
2) It is easy to understand that $\boldsymbol{B}_K^\text{T}\boldsymbol{B}_K$ is a full rank Hermitian matrix. Then, according to the conclusion in \cite{denton2021eigenvectors}, the element square values in the eigenvector matrix $\boldsymbol{D}$ can be calculated by the eigenvalues of $\boldsymbol{B}_K^\text{T}\boldsymbol{B}_K$ and its minor matrices. Eq. (\ref{eq:A2}) shows how to calculate the first row elements in $\boldsymbol{D}$ by this way.
\begin{align}
    |d_{1,i}|^2=\frac{\prod\limits_{j=1}^{K}(\delta_i(\boldsymbol{B}_K^\text{T}\boldsymbol{B}_K)-\delta_j(\boldsymbol{W}_{11}))}{\prod\limits_{j=1,j\neq i}^{K+1}(\delta_i(\boldsymbol{B}_K^\text{T}\boldsymbol{B}_K)-\delta_j(\boldsymbol{B}_K^\text{T}\boldsymbol{B}_K))},1\le i \le (K+1),
    \label{eq:A2}
\end{align}
where $\boldsymbol{W}_{11}$ is a minor obtained by deleting the first row and column elements of $\boldsymbol{B}_K^\text{T}\boldsymbol{B}_K$ in (\ref{eq:A3}); $\delta$ is the eigenvalue. As shown in (\ref{eq:A2}), the first row elements in $\boldsymbol{D}$ are closely related to the eigenvalues of matrices $\boldsymbol{B}_K^\text{T}\boldsymbol{B}_K$ and $\boldsymbol{W}_{11}$. The elements of these two matrices are shown in (\ref{eq:A4}).
\begin{align}
    \boldsymbol{B}_K^\text{T}\boldsymbol{B}_K=
    \begin{pmatrix}
     N & 0 & 2\sum\limits_{n=1}^{N_h}\frac{(nT_\text{s})^2}{2!}  &  \ldots \\
     0 & 2\sum\limits_{n=1}^{N_h}(nT_\text{s})^2 & 0 &  \ldots\\
     2\sum\limits_{n=1}^{N_h}\frac{(nT_\text{s})^2}{2!} & 0 &2\sum\limits_{n=1}^{N_h}\frac{(nT_\text{s})^4}{2!\cdot2!} &  \ldots\\
     \vdots & \vdots & \vdots & \ddots 
    \end{pmatrix}.
    \label{eq:A3}
\end{align}
\begin{numcases}{\underset{1\le i,\,j \le (K+1)}{(\boldsymbol{B}_K^\text{T}\boldsymbol{B}_K)_{ij}}=}
0, & $(i+j)$ is odd,\nonumber\\
N, & $i=j=1$,\nonumber\\
2\sum\limits_{n=1}^{N_h}\frac{(nT_\text{s})^{(i+j-2)}}{(i-1)!\cdot (j-1)!}, & $(i+j)$ is even,\nonumber
\end{numcases}
\begin{numcases}{\underset{1\le i,\,j \le K}{(\boldsymbol{W}_{11})_{ij}}=}
0, & $(i+j)$ is odd,\nonumber\\
2\sum\limits_{n=1}^{N_h}\frac{(nT_\text{s})^{(i+j)}}{i!\cdot j!}, & $(i+j)$ is even.
\label{eq:A4}
\end{numcases}

Eq. (\ref{eq:A4}) shows that $\boldsymbol{B}_K^\text{T}\boldsymbol{B}_K$ and $\boldsymbol{W}_{11}$ are formed in a similar form. As a result, their eigenvalues have a regular pattern as $\delta_{(2a)}(\boldsymbol{B}_K^\text{T}\boldsymbol{B}_K)=\delta_{(2a-1)}(\boldsymbol{W}_{11})\enskip(1\le a\le\lfloor(K+1)/2\rfloor)$. Then, the mentioned finding can be proved by (\ref{eq:A5}).
\begin{align}
     |d_{1(2a+1)}|^2&=\frac{\prod\limits_{j=1}^{K}(\delta_{(2a+1)}(\boldsymbol{B}_K^\text{T}\boldsymbol{B}_K)-\delta_j(\boldsymbol{W}_{11}))}{\prod\limits_{j=1,j\neq (2a+1)}^{K+1}(\delta_{(2a+1)}(\boldsymbol{B}_K^\text{T}\boldsymbol{B}_K)-\delta_j(\boldsymbol{B}_K^\text{T}\boldsymbol{B}_K))}\nonumber\\
   &\neq0,\quad0\le a\le\lfloor K/2\rfloor.
    \nonumber\
    \nonumber\\
    |d_{1(2a)}|^2&=(\delta_{(2a)}(\boldsymbol{B}_K^\text{T}\boldsymbol{B}_K)-\delta_{(2a-1)}(\boldsymbol{W}_{11}))\times\nonumber\\
    &\quad\quad\frac{\prod\limits_{j=1,j\neq(2a-1)}^{K}(\delta_{(2a)}(\boldsymbol{B}_K^\text{T}\boldsymbol{B}_K)-\delta_j(\boldsymbol{W}_{11}))}{\prod\limits_{j=1,j\neq (2a)}^{K+1}(\delta_{(2a)}(\boldsymbol{B}_K^\text{T}\boldsymbol{B}_K)-\delta_j(\boldsymbol{B}_K^\text{T}\boldsymbol{B}_K))}\nonumber\\
    &=0,\quad1\le a\le\lfloor(K+1)/2\rfloor.
    \label{eq:A5}
\end{align}

\bibliographystyle{IEEEtran}

\end{document}